\documentclass[12pt]{article}
\pdfoutput=1

\usepackage{mathrsfs}
\usepackage[T1]{fontenc}
\usepackage{mathpazo}
\usepackage{setspace}
\usepackage{amsfonts}
\usepackage{amssymb}
\usepackage{amsmath}
\usepackage{esint}                   
\usepackage{epsfig}
\usepackage{float}
\usepackage{latexsym}
\usepackage{color}
\usepackage{graphicx}
\usepackage{hyperref}
\usepackage{nicefrac}
\usepackage[latin1]{inputenc}
\usepackage{pstricks}
\usepackage{slashed}
\usepackage{multirow}
\usepackage{cancel}
\usepackage{empheq}
\usepackage{pifont}
\usepackage{caption}
\usepackage{subcaption}
\usepackage[textwidth= 18cm, height= 23cm]{geometry}

\def\baselinestretch{1.2}

\catcode`\@=11

\def\marginnote#1{}
\newcount\hour
\newcount\minute
\newtoks\amorpm
\hour=\time\divide\hour by60
\minute=\time{\multiply\hour by60 \global\advance\minute by-\hour}
\edef\standardtime{{\ifnum\hour<12 \global\amorpm={am}%
        \else\global\amorpm={pm}\advance\hour by-12 \fi
        \ifnum\hour=0 \hour=12 \fi
        \number\hour:\ifnum\minute<10 0\fi\number\minute\the\amorpm}}
\edef\militarytime{\number\hour:\ifnum\minute<10 0\fi\number\minute}

\def\draftlabel#1{{\@bsphack\if@filesw {\let\thepage\relax
   \xdef\@gtempa{\write\@auxout{\string
      \newlabel{#1}{{\@currentlabel}{\thepage}}}}}\@gtempa
   \if@nobreak \ifvmode\nobreak\fi\fi\fi\@esphack}
        \gdef\@eqnlabel{#1}}
\def\@eqnlabel{}
\def\@vacuum{}
\def\draftmarginnote#1{\marginpar{\raggedright\scriptsize\tt#1}}

\def\draft{\oddsidemargin -.5truein
        \def\@oddfoot{\sl preliminary draft \hfil
        \rm\thepage\hfil\sl\today\quad\militarytime}
        \let\@evenfoot\@oddfoot \overfullrule 3pt
        \let\label=\draftlabel
        \let\marginnote=\draftmarginnote
   \def\@eqnnum{(\theequation)\rlap{\kern\marginparsep\tt\@eqnlabel}%
\global\let\@eqnlabel\@vacuum}  }

\def\draft2{
        \def\@oddfoot{\sl preliminary draft \hfil
        \rm\thepage\hfil\sl\today\quad\militarytime}
        \let\@evenfoot\@oddfoot \overfullrule 3pt
        \let\label=\draftlabel
        \let\marginnote=\draftmarginnote
   \def\@eqnnum{(\theequation)\rlap{\kern\marginparsep\tt\@eqnlabel}%
\global\let\@eqnlabel\@vacuum}  }

\def\preprint{\twocolumn\sloppy\flushbottom\parindent 2em
        \leftmargini 2em\leftmarginv .5em\leftmarginvi .5em
        \oddsidemargin -.5in    \evensidemargin -.5in
        \columnsep .4in \footheight 0pt
        \textwidth 10.in        \topmargin  -.4in
        \headheight 12pt \topskip .4in
        \textheight 6.9in \footskip 0pt
        \def\@oddhead{\thepage\hfil\addtocounter{page}{1}\thepage}
        \let\@evenhead\@oddhead \def\@oddfoot{} \def\@evenfoot{} }

\def\numberbysection{\@addtoreset{equation}{section}
        \def\theequation{\thesection.\arabic{equation}}}

\def\underline#1{\relax\ifmmode\@@underline#1\else
        $\@@underline{\hbox{#1}}$\relax\fi}

\def\titlepage{\@restonecolfalse\if@twocolumn\@restonecoltrue\onecolumn
     \else \newpage \fi \thispagestyle{empty}\c@page\z@
        \def\thefootnote{\fnsymbol{footnote}} }

\def\endtitlepage{\if@restonecol\twocolumn \else \newpage \fi
        \def\thefootnote{\arabic{footnote}}
        \setcounter{footnote}{0}}  

\catcode`@=12
\relax

\def\figcap{\section*{Figure Captions\markboth
        {FIGURECAPTIONS}{FIGURECAPTIONS}}\list
        {Figure \arabic{enumi}:\hfill}{\settowidth\labelwidth{Figure
999:}
        \leftmargin\labelwidth
        \advance\leftmargin\labelsep\usecounter{enumi}}}
 \relax
\def\tablecap{\section*{Table Captions\markboth
        {TABLECAPTIONS}{TABLECAPTIONS}}\list
        {Table \arabic{enumi}:\hfill}{\settowidth\labelwidth{Table
999:}
        \leftmargin\labelwidth
        \advance\leftmargin\labelsep\usecounter{enumi}}}
 \relax
\def\reflist{\section*{References\markboth
        {REFLIST}{REFLIST}}\list
        {[\arabic{enumi}]\hfill}{\settowidth\labelwidth{[999]}
        \leftmargin\labelwidth
        \advance\leftmargin\labelsep\usecounter{enumi}}}
 \relax
\makeatletter
\newcounter{pubctr}
\def\publist{\@ifnextchar[{\@publist}{\@@publist}}
\def\@publist[#1]{\list
        {[\arabic{pubctr}]\hfill}{\settowidth\labelwidth{[999]}
        \leftmargin\labelwidth
        \advance\leftmargin\labelsep
        \@nmbrlisttrue\def\@listctr{pubctr}
        \setcounter{pubctr}{#1}\addtocounter{pubctr}{-1}}}
\def\@@publist{\list
        {[\arabic{pubctr}]\hfill}{\settowidth\labelwidth{[999]}
        \leftmargin\labelwidth
        \advance\leftmargin\labelsep
        \@nmbrlisttrue\def\@listctr{pubctr}}}
 \relax
\makeatother

\def\be{\begin{equation}}
\def\ee{\end{equation}}
\def\ba{\begin{eqnarray}}
\def\ea{\end{eqnarray}}

\def\vol{\textrm{Vol}}

\def\r{\rho}
\def\a{\alpha}

\def\b{\beta}

\def\g{\gamma}
\def\G{\Gamma}
\def\d{\delta}
\def\D{\Delta}

\def\m{\mu}
\def\n{\nu}

\def\Om{\Omega}
\def\l{\lambda}

\def\s{\sigma}

\def\cA{{\cal A}}

\def\cC{{\cal C}}
\def\cD{{\cal D}}

\def\cL{{\cal L}}

\def\cM{{\cal M}}
\def\cN{{\cal N}}
\def\cO{{\cal O}}

\def\cY{{\cal Y}}

\def\no{\noindent}

\def\IR{\relax{\rm I\kern-.18em R}}

\def\inv{^{\raise.0ex\hbox{${\scriptscriptstyle -}$}\kern-.05em 1}}

\begin{document}


\renewcommand{\theequation}{\thesection.\arabic{equation}}
\csname @addtoreset\endcsname{equation}{section}

\begin{titlepage}
\begin{center}

\phantom{xx}
\vskip 0.5in

{\large \bf Spin-2 excitations in Gaiotto-Maldacena solutions}

\vskip 0.5in

{\bf Georgios Itsios}${}^{1a},$\phantom{x} {\bf Jos\'e Manuel Pen\'in}${}^{2,3b}$\phantom{x} {\bf and \phantom{x} Salom\'on Zacar\'ias}${}^{4c}$ \vskip 0.1in

${}^1$ Instituto de F\'isica Te\'orica, UNESP-Universidade Estadual Paulista, \\
R. Dr. Bento T. Ferraz 271, Bl. II, Sao Paulo 01140-070, SP, Brazil\\

\vskip .2in

${}^2$ Departamento de F\'isica de Part\'iculas and\\${}^3$ Instituto Galego de F\'isica de Altas Enerx\'ias (IGFAE)\\Universidade de Santiago de Compostela\\
 E-15782 Santiago de Compostela, Spain
 
 \vskip .2in
 
 ${}^4$ Shanghai Center for Complex Physics, Department of Physics and Astronomy, \\
 Shanghai JiaoTong University, Shanghai 200240, China.

\end{center}

\vskip .4in

\centerline{\bf Abstract}

\no

In this paper we study spin-2 excitations for a class of $\cN = 2$ supersymmetric solutions of type-IIA supergravity found by Gaiotto and Maldacena. The mass spectrum of these excitations can be derived by solving a second order partial differential equation. As specific examples of this class we consider the Abelian and non-Abelian T-dual versions of the $AdS_5 \times S^5$ and we study the corresponding mass spectra. For the modes that do not ``feel'' the (non-)Abelian T-duality transformation we provide analytic formulas for the masses, while for the rest we were only able to derive the spectra numerically. The numerical values that correspond to large masses are compared with WKB approximate formulas. We also find a lower bound for the masses. Finally, we study the field theoretical implications of our results and propose dual spin-2 operators.

\vfill
\no
 {
$^a$gitsios@gmail.com,\\  
$^b$jmanpen@gmail.com\\
$^c$szacarias@sjtu.edu.cn
}

\end{titlepage}
\vfill
\eject



\def\baselinestretch{1.2}
\baselineskip 20 pt

\newcommand{\eqn}[1]{(\ref{#1})}

\tableofcontents

\section{Introduction and summary of results}

In the last two decades there have been significant developments towards the understanding of the strongly coupled dynamics of supersymmetric quantum field theories in four dimensions.
In this paper we will study $\mathcal{N}=2$ supersymmetric theories using holography \cite{Maldacena:1997re}. One can understand these theories as arising from stacks of $NS5$ and $D6$ 
branes with $D4$ branes stretching and intersecting them. At low energies the world-volume theory of the $D4-NS5-D6$ brane arrangement is described by linear quiver gauge theories with product $SU(N_c)$ 
gauge groups connected by bi-fundamental fields and $SU(N_f)$ fundamental matter for each gauge group associated to the $D6$-branes. In the conformal case these theories preserve $SO(4,2)\times SU(2)\times U(1)$ 
isometries and arise as a compactification of $M5$ branes on a punctured Riemann surface \cite{Gaiotto:2009we}.

 Gravitational solutions holographically dual to these theories were found by Gaiotto and Maldacena in  \cite{Gaiotto:2009gz}. Based on the previous work by LLM \cite{Lin:2004nb},  it was found that by 
 considering a smearing of the M5 branes along the eleven dimensional coordinate one can reduce these solutions to type-IIA supergravity. This generates a whole class of solutions in terms of a function
 solving an axisymmetric Laplace equation with appropriate boundary conditions that ensure regularity of the solution and proper quantization of $D$-brane fluxes. Given a solution 
 one can construct the dual quiver field theory with the rules spelled out in \cite{Gaiotto:2009gz} (see also  \cite{Aharony:2012tz}). One can then use holography to compute observables and learn about the $\mathcal{N}=2$ 
 theories at strong coupling. Recently, using this approach, several new formulas computing field theory observables in terms of geometric data were presented in \cite{Nunez:2019gbg}. 
 Generic solutions for this class of geometries and some particular interesting examples were studied in  \cite{Aharony:2012tz,ReidEdwards:2010qs,Petropoulos:2014rva,Petropoulos:2013vya}. Finding new examples 
  belonging to this class of geometries however is challenging and solution generating techniques like non-Abelian T-duality have proven to be very useful to construct new examples \cite{Sfetsos:2010uq}.

In this paper we obtain the linearized equations of motion for the fluctuations of the type-IIA supergravity fields around an arbitrary background.
We restrict our attention to a consistent truncation, namely that of spin-2 excitations of the $\mathcal{N}=2$ geometries discussed above. As a result we have to deal with a second order differential equation which coincides with the analysis in \cite{Bachas:2011xa} (see also \cite{Passias:2018swc,Passias:2016fkm,Richard:2014qsa,Gutperle:2018wuk,Chen:2019ydk} for similar studies of the spin-2 excitations). We provide a generic expression 
for the wave operator given in terms of the function that solves the axisymmetric Laplace equation. We use the above operator to study the spectrum of two interesting examples. The first of them is the Abelian (Hopf) T-dual (ATD) of the $AdS_5\times S^5$ solution \cite{Fayyazuddin:1999zu,Lozano:2016kum}. Regardless this solution does not satisfy the appropriate boundary conditions for the axisymmetric Laplacian, it is still a good solution. The second example is 
the  Sfetsos-Thompson solution \cite{Sfetsos:2010uq} obtained after applying non-Abelian T-duality (NATD) along the $SU(2)$ isometries inside the $S^5$ of the maximally supersymmetric solution 
in the type-IIB supergravity. It turns out that this solution defines a singular geometry. Despite this issue, this solution has some interesting properties that make it stand out from others belonging to this 
class of geometries. For instance, it was shown to be an integrable background \cite{Nunez:2018qcj} as opposed to the  generic ``smooth'' non-singular solutions of the large class 
of Gaiotto-Maldacena geometries \cite{Borsato:2017qsx,Wulff:2019tzh}. A detailed study of the field theory dual of the Sfetsos-Thompson solution including a completion to the geometry can be found in  \cite{Lozano:2016kum}.
It is worth noticing that the ATD solution can be obtained as a limiting case of the NATD one\footnote{Though the dilaton and the RR fields did not match, this issue was solved in \cite{vanGorsel:2017goj}.} 
 \cite{Lozano:2016kum}. We also see this relation at the level of the operators describing the spin-2 excitations. 
Using holography we can then shed some 
 light towards the understanding of the operator spectrum of the dualized solutions. In the BMN limit this problem was studied in \cite{Itsios:2017nou}.
 
In order to analyze the spin-2 spectrum of the solutions, we transformed the equation for the fluctuations into a Schr\"{o}dinger-like problem of a particle in certain potential. The resulting potential is very similar 
to the one considered in the study of marginal deformations of supersymmetric backgrounds  \cite{Hernandez:2005xd,Hernandez:2005zx}. In both ATD and NATD examples we were able to find analytic solutions
only for the zero value of a quantum number $n$. For non-zero values of this quantum number we solved the equation numerically using the shooting method. For large masses we compared these solutions with a WKB 
analysis finding agreement. The analysis of the NATD solution shows a continuous spectrum of masses. This issue is associated with the fact that the ``field space'' coordinate $\rho$  in the solution is unbounded. 
A discrete spectrum arises whenever we bound the value of this coordinate. This hard cut-off in the geometry is ascertained by placing $D6$ branes at a certain position $\rho_{*}$  \cite{Lozano:2016kum}. The situation with the
ATD solution is very similar, though in this case the ``field space'' coordinate is periodic giving rise to a discrete spectrum. 

The analysis of the spectrum of the above discussed examples gives states dual to spin-2 operators of 4d $\mathcal{N}=2$ SCFTs described by linear quivers involving vector multiplets as well as hypermultiplets in the bifundamental representation. 
Such operators have dimension $\Delta=4+2\ell+\vert m\vert+\nu$. We use the field content of the dual SCFTs in order to define an intuitive structure of these spin-2 operators with the appropriate dimension and R-charges.

 The paper is organized as follows. In section 2 we briefly review the Gaiotto-Maldacena class of geometries. In section 3 we study the spin-2 excitations of these geometries by perturbing the metric components along the $AdS_5$ sector.
 We write down a generic expression for the operator describing the fluctuations in terms of the function that solves the axisymmetric Laplace equation. In section 4 we study the spectrum of two particular examples of 
Gaiotto-Maldacena geometries: the ATD and NATD solutions. In both cases we performed analytical as well as numerical methods to obtain the 
spectrum and its lower bound. In section 5 we discuss the field theory interpretation of our results. We conclude in section 6 with a brief summary of our results and future directions. We provide detailed appendices with the machinery needed to 
present the results of this paper. Appendix A is a compendium of formulas that are useful to derive the fluctuation equations. Appendix B contains the equations for the fluctuations of all the fields in type-IIA supergravity
 in the Einstein frame. Finally in appendix C we discuss the WKB approach that we used to give support to our numerical study of the spectrum in section 4.

{\it \underline{Note added}:} When we were finishing this paper we became aware of \cite{Chen:2019ydk} which has some overlap with the content of our section 3.

\section{Gaiotto-Maldacena solutions}

The theories that we are going to deal with here are $\cN = 2$ supersymmetric solutions of the type-IIA supergravity, which have been found in \cite{Gaiotto:2009gz} and whose metric in the \emph{Einstein frame} has the following form:
\begin{equation}
\label{GMmetric}
 ds^2_E = e^{-\frac{\Phi}{2}} f_0 \, \Big(  ds^2_{AdS_5} + f_1 d\Om^2_2 + f_2 \big(  d\eta^2 + d\s^2 \big) + f_3 d\b^2  \Big) \, ,
\end{equation}
with $d\Om^2_2 = d\chi^2 + \sin^2\chi d\xi^2$ being the line element of a two-sphere. The dilaton $\Phi$ and the functions $f_i$ depend only on the coordinates $(\eta, \s)$ and they can be expressed in terms of a function $V(\eta, \s)$ as:
\begin{equation}
 \begin{aligned}\label{functionsv}
  & e^{2 \Phi} = 2^7 \s \frac{\big(  \ddot{V} - 2 \, \dot{V} \big)^{3/2}}{\sqrt{\ddot{V}} \dot{V} \D} \, , \qquad \D = \big(  \ddot{V} - 2 \dot{V} \big) \frac{\ddot{V}}{\s^2} + \dot{V}'^2 \, , \qquad f_0 = \s \sqrt{\frac{\ddot{V} - 2 \, \dot{V}}{\ddot{V}}} \, ,
  \\[5pt]
  & f_1 = - \frac{\ddot{V} \dot{V}}{2 \, \s^2 \D} \, ,\qquad f_2 = - \frac{\ddot{V}}{2 \,\s^2 \dot{V}} \, , \qquad f_3 = \frac{\ddot{V}}{\ddot{V} - 2 \, \dot{V}} \, .
 \end{aligned}
\end{equation}
Notice that due to the normalizations we adopted here the $AdS_5$ space has a unit radius. Moreover, primed symbols correspond to derivatives with respect to $\eta$ while dotted symbols correspond to the action of the operator $\s \partial_\s$.

The geometry of this class of solutions is supported by a NS two-form and a set of RR potentials:
\begin{equation}
 B_2 = \frac{1}{2} \Big(  \frac{\dot{V} \dot{V}'}{\D} - \eta \Big) \vol_{\Om_2} \, , \qquad
 C_1 = \frac{1}{8} \, \frac{\dot{V} \dot{V}'}{2 \dot{V} - \ddot{V}} d\b \, , \qquad
 C_3 = \frac{1}{16} \, \frac{\dot{V}^2 \ddot{V}}{\s^2 \D} d\b \wedge \vol_{\Om_2} \, ,
\end{equation}
where $\vol_{\Om_2} = \sin\chi d\chi \wedge d\xi$ is the volume form on the two-sphere $\Om_2$ and the RR fields $F_2$ and $F_4$ are defined through the potentials $C_1$ and $C_3$ as $F_2 = dC_1$ and $F_4 = dC_3 + C_1 \wedge H_3$ with $H_3 = d B_2$.

As it is understood by the previous expressions, any background that fits into the Gaiotto-Maldacena classification is fully determined by the function $V(\eta, \s)$. However this function is not arbitrary but instead it has to satisfy the following second order differential equation:
\begin{equation}
\label{condV}
 \ddot{V} + \s^2 V'' = 0 \, .
\end{equation}

In addition, the above equation is supplemented with boundary conditions for the function $V(\sigma,\eta)$ and its derivatives that fully determine the solution and ensure the geometry and matter fields obtained via (\ref{functionsv}) are 
regular and properly quantised. We requiere $V(\sigma\rightarrow \infty,\eta)\rightarrow0$. In addition, by defining  the charge density $\lambda(\eta)$ the remaining boundary conditions read
\begin{equation}
\label{condV2}
\dot{V}(\sigma,\eta)\vert_{\sigma=0}=\lambda(\eta),\quad \lambda(\eta)>0,\quad \lambda(0)=\lambda(\eta_{*})=0.
\end{equation}
The above function is also required to be piece-wise continuous, made out of linear segments $\lambda(\eta)=a_i\eta+b_i$,  $a_i\in \mathbb{Z}$ with $a_i-a_{i-1}<0$. The change in slope $a_{i}$ between consecutive segments must be in addition 
an integer and is related to the presence of $D6$ branes in the geometry. 
Therefore a background constructed by  a given $\dot{V}$ satisfying the above conditions is dual to 4d $\mathcal{N}=2$ SCFTs of the GM class.

In the section that follows we will study perturbations of the metric \eqref{GMmetric} along the $AdS_5$ directions.

\section{Metric perturbations}

In this section we look for a consistent truncation of the equations for the fluctuation of the supergravity fields. Before we start this analysis it is worth to mention some properties of the geometry of the solutions that we are interested in. The first property is that the metric in the \emph{Einstein frame} is conformal to a direct sum of two five-dimensional spaces. More specifically it is conformal to the sum of $AdS_5$ with a five-dimensional internal space $\cM_5$:
\begin{equation}
\label{metricConf}
 ds^2 = ds^2(AdS_5) + ds^2(\cM_5) \, .
\end{equation}
Thus, it is useful to adopt the following notation for the indices:
\begin{equation*}
 \begin{array}{ll}
    M,N,P,K,\Lambda,\Sigma, \ldots : \qquad &\textrm{ten-dimensional indices} ,
    \\[5pt]
    \mu,\nu,\rho,\kappa,\lambda,\s, \ldots : \qquad &\textrm{indices in } AdS_5 ,
    \\[5pt]
    m,n,k,p,q,s, \ldots : &\textrm{indices in } \cM_5 .
 \end{array}
\end{equation*}
Moreover we will consider the splitting of the coordinates $X^M = (x^\mu, y^m)$, where $x$ are the coordinates in $AdS_5$ and $y$ the coordinates in $\cM_5$. The ten-dimensional line element \eqref{metricConf} can be written as:
\begin{equation}
 \tilde{g}_{MN} \, dX^M \, dX^N = \tilde{g}_{\mu\nu} \, dx^{\mu} \, dx^{\nu} + \tilde{g}_{mn} \, dy^m \, dy^n \, ,
\end{equation}
where the metric components $\tilde{g}_{\mu\nu}$ depend only on $x$ and $\tilde{g}_{mn}$ only on $y$. Using matrix notation the metric reads:
\begin{equation}
 \tilde{g}_{MN}(x,y) = \begin{pmatrix}
 			          \tilde{g}_{\mu\nu}(x) & 0
			         \\[5pt]
			          0 & \tilde{g}_{mn}(y)
 			         \end{pmatrix} \, .
\end{equation}
The second important property of the geometries that we are going to consider is that the metric is diagonal.

Next we would like to turn on only the fluctuations of the metric components along the $AdS_5$ sector, while the fluctuations of the rest of the fields are taken to be zero, i.e. we take into account only the following:
\begin{equation}
 \d g_{\m\nu} = e^{2 A} h_{\mu\nu} \, ,
\end{equation}
or
\begin{equation}
 ds^2_E = e^{2A} \Big[  \big(  \tilde{g}_{\mu\nu} + h_{\mu\nu} \big) dx^{\mu} dx^{\nu} + \tilde{g}_{mn} dy^m dy^n  \Big] \, .
\end{equation}
In addition we take $h_{\mu\nu}$ to factorize as:
\begin{equation}
 h_{\mu\nu}(x,y) = h^{[tt]}_{\mu\nu}(x) Y(y) \, ,
\end{equation}
where $h^{[tt]}_{\mu\nu}$ is transverse with respect to $\tilde{\nabla}^{\mu}$ and traceless, i.e.
\begin{equation}
 \tilde{\nabla}^{\mu} h^{[tt]}_{\mu\nu} = 0 \, , \qquad \tilde{g}^{\mu\nu} h^{[tt]}_{\mu\nu} = 0 \, .
\end{equation}
Under these considerations we see that the fluctuation of the dilaton equation and also those for the Maxwell equations are trivially satisfied. However from the Einstein equations we see that the only terms that contribute are:
\begin{equation}
 \begin{aligned}
  0 & = \frac{1}{2} \, \tilde{\nabla}^{\Sigma} \tilde{\nabla}_M h_{\Sigma N} + \frac{1}{2} \, \tilde{\nabla}^{\Sigma} \tilde{\nabla}_N h_{\Sigma M} - \frac{1}{2} \, \tilde{\nabla}^2 h_{MN}
  - 4 \, \tilde{\nabla}^P A \tilde{\nabla}_P h_{MN} - h_{MN} \,  \tilde{\nabla}^2 A - 8 \, h_{MN} \, \big(  \tilde{\nabla} A  \big)^2
  \\[5pt]
  & + \frac{1}{2} h_{MN} \sum\limits_{p = 2}^{4} \b_p \g_p \, e^{2 (1 - p) A + \a_p \bar{\Phi}} \tilde{\cA}^2_p \, ,
 \end{aligned}
\end{equation}
where the notation of $\tilde{\cA}^2_p$ is explained in appendix \ref{appEinsteinEq}. In order to simplify the above expression we took into account the structure of the background fields (coordinate dependence and index structure) of the Gaiotto-Maldacena solutions. This can be further simplified if we change the order of the covariant derivatives of the first two terms. Using \ref{ComCD}, we get:
\begin{equation}
 \begin{aligned}
  \tilde{\nabla}^{\Sigma}  \tilde{\nabla}_M h_{\Sigma N} & = \tilde{\nabla}_M  \tilde{\nabla}^{\Sigma} h_{\Sigma N} + \tilde{g}^{P P'} \, \tilde{R}_{P' M} \; h_{P N} - \tilde{g}^{K \Sigma} \, \tilde{R}^{P}_{\;\; NKM} \; h_{\Sigma P} \, .
 \end{aligned}
\end{equation}
The first term vanishes due to the transversality condition. Now the Einstein equation becomes:
\begin{equation}
 \begin{aligned}
  0 & = \tilde{g}^{\r \s} \, \tilde{R}_{\s \mu} \; h_{\r \nu} - \tilde{g}^{\kappa \s} \, \tilde{R}^{\r}_{\;\; \nu \kappa \mu} \; h_{\s \r}
  + \tilde{g}^{\r \s} \, \tilde{R}_{\s \nu} \; h_{\r \mu} - \tilde{g}^{\kappa \s} \, \tilde{R}^{\r}_{\;\; \mu \kappa \nu} \; h_{\s \r}
   - \tilde{\nabla}^2 h_{\mu \nu}
  \\[5pt]
  & - 8 \, \tilde{\nabla}^P A \tilde{\nabla}_P h_{\mu \nu} - 2 \,h_{\mu \nu} \,  \tilde{\nabla}^2 A - 16 \, h_{\mu \nu} \, \big(  \tilde{\nabla} A  \big)^2
  + h_{\mu \nu} \sum\limits_{p = 2}^{4} \b_p \g_p \, e^{2 (1 - p) A + \a_p \bar{\Phi}} \tilde{\cA}^2_p \, .
 \end{aligned}
\end{equation}
We recall that the Riemann and Ricci tensors of the $AdS_5$ of unit radius are:
\begin{equation}
 \tilde{R}_{\mu\nu\rho\s} = \tilde{g}_{\mu\s} \tilde{g}_{\nu\rho} - \tilde{g}_{\mu\rho} \tilde{g}_{\nu\s} \qquad \Rightarrow \qquad \tilde{R}_{\nu\s} = - 4 \, \tilde{g}_{\nu \s} \, .
\end{equation}
So
\begin{equation}
   \tilde{g}^{\r \s} \, \tilde{R}_{\s \mu} \; h_{\r \nu} = - 4 \; h_{\mu \nu} \, ,
\qquad
   \tilde{g}^{\kappa \s} \, \tilde{R}^{\r}_{\;\; \nu \kappa \mu} \; h_{\s \r} = h_{\mu \nu} \, .
\end{equation}
Therefore, the equation that we need to solve is:
\begin{equation}
  0 = \tilde{\nabla}^2 h_{\mu \nu} + 10 \; h_{\mu \nu}
  + 8 \, \tilde{\nabla}^P A \tilde{\nabla}_P h_{\mu \nu} + h_{\mu \nu} \, \Big[ 2 \, \tilde{\nabla}^2 A + 16 \, \big(  \tilde{\nabla} A  \big)^2
  - \sum\limits_{p = 2}^{4} \b_p \g_p \, e^{2 (1 - p) A + \a_p \bar{\Phi}} \tilde{\cA}^2_p \Big] \, .
\end{equation}
It turns out that, for all the solutions that belong to the Gaiotto-Maldacena class, the term in square brackets equals to $-8$, thus the equation that we have to solve is:
\begin{equation}
\label{EinsteinReduced}
  0 = \tilde{\nabla}^{\s} \tilde{\nabla}_{\s} h_{\mu \nu} + 2 \; h_{\mu \nu}
  + \tilde{\nabla}^{m} \tilde{\nabla}_{m} h_{\mu \nu} + 8 \, \tilde{\nabla}^m A \tilde{\nabla}_m h_{\mu \nu} \, .
\end{equation}
Notice that the last two terms can be written as:
\begin{equation}
 \tilde{\nabla}^{m} \tilde{\nabla}_{m} h_{\mu \nu} + 8 \, \tilde{\nabla}^m A \tilde{\nabla}_m h_{\mu \nu} = e^{- 8A} \tilde{\nabla}^m \Big[  e^{8 A} \tilde{\nabla}_m h_{\mu \nu} \Big] := \cL ( h_{\mu \nu} ) \, .
\end{equation}
This has exactly the same form as the operator $\cL^{(1)}$ in \cite{Passias:2018swc}. Also, since the indices of $h_{\mu \nu}$ are along the $AdS_5$ subspace we understand that $h_{\mu \nu}$ behaves like a scalar for the covariant derivative $\tilde{\nabla}_m$ and the operator $\cL$. Now the action of $\cL$ on a scalar $f$ can be written as:
\begin{equation}
 \cL(f) = \tilde{\nabla}^m \tilde{\nabla}_m f + 8 \; \tilde{\nabla}^m A \; \tilde{\nabla}_m f = \frac{1}{\sqrt{\tilde{g}_{\cM_5}}} \partial_m \Big(  \sqrt{\tilde{g}_{\cM_5}} \; \tilde{g}^{mn} \; \partial_n f  \Big) + 8 \; \tilde{g}^{mn} \; \partial_m A \; \partial_n f \; .
\end{equation}
Moreover, the equation \eqref{EinsteinReduced} can be recognized as the equation of motion of a massive graviton of mass $M$ propagating in $AdS_5$ \cite{Polishchuk:1999nh,Buchbinder:1999be}. This is given by the Pauli-Fierz equation:
\begin{equation}
0 =  \tilde{\nabla}^{\s} \tilde{\nabla}_{\s} h_{\mu \nu} + \big(  2 - M^2  \big) \; h_{\mu \nu} \; .
\end{equation}
Using this, the equation \eqref{EinsteinReduced} reduces to an eigenvalue problem for the operator $\cL$:
\begin{equation}
\label{EigenvalueProblem}
 \cL (h_{\mu \nu}) = - M^2 h_{\mu\nu} \, .
\end{equation}
In terms of the coordinates of the metric \eqref{GMmetric} the operator $\cL$ has the following form:
\begin{equation}
 \cL (f) = \frac{1}{f_1} \nabla^2_{(2)} f + \frac{1}{f_3} \partial^2_{\b} f + \frac{1}{\D f_0 f_1 \sqrt{f_3}} \Bigg[  \partial_{\eta} \Big(  \frac{\D f_0 f_1 \sqrt{f_3}}{f_2} \partial_{\eta} f \Big) + \partial_{\s} \Big(  \frac{\D f_0 f_1 \sqrt{f_3}}{f_2} \partial_{\s} f \Big) \Bigg] \, ,
\end{equation}
where $\nabla^2_{(2)}$ is the Laplace operator on the two-sphere $\Om_2$. This operator looks quite complicated and for this reason finding its eigenvalues for the general case of an arbitrary solution of \eqref{condV} is a non-trivial task. However, as we will immediately see in the following section, one can focus on specific solutions of \eqref{condV} which lead to a solvable eigenvalue problem for the operator $\cL$.

\section{The spin-2 spectrum}

It is known \cite{Sfetsos:2010uq,Lozano:2016kum} that the T-dual of $AdS_5 \times S^5$ as well as its non-Abelian T-dual version are both examples of Gaiotto-Maldacena backgrounds. This fact motivates us to look for solutions of the eigenvalue problem \eqref{EigenvalueProblem} in the aforementioned two cases. This is feasible because as we will see shortly the operator $\cL$ simplifies significantly in both examples. Let us see this in more detail.

\subsubsection*{Operator $\cL$ in the ATD}

The Abelian T-dual solution is derived by the Gaiotto-Maldacena ansatz by choosing the potential $V(\eta, \s)$ to be \cite{Lozano:2016kum}:
\begin{equation}
 V^{ATD}(\eta, \s) = \ln \s - \frac{\s^2}{2} + \eta^2 \, .
\end{equation}
One has also to perform a change of coordinates in the following manner
\footnote{
In this notation $\psi$ is the coordinate associated to the T-duality transformation.
}:
\begin{equation}
\label{CoordChATD}
 \eta = 2 \psi \, , \qquad \s = \sin\a \, .
\end{equation}
As a result, we end up with a simple expression for the operator $\cL$:
\begin{equation}
  \cL^{ATD}(f) = \partial^2_{\a} f + \big(  \cot\a - 3 \; \tan\a \big) \partial_{\a} f + \frac{1}{\sin^2\a} \partial^2_{\b} f + \frac{\cos^2\a}{4} \; \partial^2_{\psi} f + \frac{4}{\cos^2\a} \nabla^2_{(2)} f \; ,
\end{equation}
where $\nabla^2_{(2)}$ is the Laplace operator on the two-sphere $\Om_2(\chi,\xi)$.

\subsubsection*{Operator $\cL$ in the NATD}

In the non-Abelian T-dual case the potential $V(\eta, \s)$ reads \cite{Sfetsos:2010uq}:
\begin{equation}
\label{VST}
 V^{NATD}(\eta, \s) = \eta \Big(\ln \s - \frac{\s^2}{2}\Big) + \frac{\eta^3}{3} \, .
\end{equation}
The change of coordinates that gives the NATD solution is the same as in eq. \eqref{CoordChATD} where now we are going to rename $\psi$ by $\r$
\footnote{
 Here the coordinate $\rho$ is a radial coordinate that together with the $\Om_2(\chi,\xi)$ results after non-Abelian T-dualizing a three-sphere inside the five-sphere of the $AdS_5 \times S^5$.
}.
The operator $\cL$ in this case is:
\begin{equation}
 \begin{aligned}
  \cL^{NATD}(f) & = \partial^2_{\a} f + \ \big(  \cot\a - 3 \; \tan\a \big) \partial_{\a} f + \frac{1}{\sin^2\a} \partial^2_{\b} f + \frac{\cos^2\a}{4} \; \Big(   \partial^2_{\r} f + \frac{2}{\r} \; \partial_{\r} f + \frac{1}{\r^2} \; \nabla^2_{(2)} f    \Big) 
  \\[5pt]
  & + \frac{4}{\cos^2\a} \nabla^2_{(2)} f \; .
 \end{aligned}
\end{equation}
This operator organizes nicely in terms of a Laplacian on the two-sphere $\Om_2 (\chi, \xi)$ and a Laplacian in spherical coordinates (terms in the parenthesis) on the three-dimensional Euclidean space parametrized by the radial coordinate $\r$ and the two-sphere $\Om_2(\chi, \xi)$. The eigenfunctions of the last are spherical Bessel functions that depend on the eigenvalue $\ell$. Notice that at $\r \gg 1$ we have $\cL^{NATD}(f) = \cL^{ATD}(f)$ with the identification $\r = \psi$.

Let us now try to solve the eigenvalue problem for the operators $\cL^{ATD}$ and $\cL^{NATD}$ starting with the Abelian T-dual example.

\subsection{The ATD case}

The form of the operator $\cL^{ATD}$ suggests that we should expand $Y(y)$ in eigenfunctions of the operators $\partial^2_{\b} , \; \partial^2_{\psi}$ and $\nabla^2_{(2)}$:
\begin{equation}
 \begin{aligned}
  & Y(y) = \sum\limits_{m,n,\ell,k} f_{m,n,\ell}(\a) e^{i (m \b + n \psi )} \cY^k_{\ell}(\chi, \xi) \, ,
  \\[5pt]
  & m, n \in \mathbb{Z} \, , \quad \ell = 0, 1, 2, \ldots \, , \quad k = - \ell, -\ell + 1 , \ldots , \ell \, ,
 \end{aligned}
\end{equation}
where $\cY^k_{\ell}$
\footnote{
Since the index $k$ does not enter anywhere else, we expect that the mass spectrum is $(2 \ell + 1)$-times degenerate.
}
are the spherical harmonics on the two-sphere $\Om_2(\chi, \xi)$ and $f_{m,n,\ell}(\a)$ are functions to be determined. Using such an expansion the eigenvalue problem \eqref{EigenvalueProblem} translates to a second order differential equation for the functions $f_{m,n,\ell}(\a)$:
\begin{equation}
\label{EVATD}
  0 = \partial^2_{\a} f + \big(  \cot\a - 3 \; \tan\a \big) \partial_{\a} f + \Bigg[   M^2 - \frac{m^2}{\sin^2\a} - \frac{n^2}{4} \cos^2\a - \frac{4 \ell (\ell+1)}{\cos^2\a}  \Bigg]  f \; ,
\end{equation}
where for convenience we have suppressed the indices $m,n,\ell$ in the function $f(\a)$. If we perform the following change of variables:
\begin{equation}
 z = \sin^2 \a \, , \qquad z \in [0,1] \, ,
\end{equation}
then the differential equation that we have to solve becomes:
\begin{equation}
\label{DEtoSolve}
  0 = z (1-z) \; \partial^2_z f + (1 - 3 z) \; \partial_z f + \Bigg[  \frac{M^2}{4} - \frac{\ell (\ell + 1)}{1 - z} - \frac{m^2}{4 \; z} - \frac{n^2}{16} (1 - z)  \Bigg] f \; .
\end{equation}
For $n = 0$ this equation can be brought into a hypergeometric form and thus it can be solved analytically. For $n \ne 0$ the above equation has two regular singular points at $z = 0, 1$ and one irregular singular point at $z = \infty$. Thus it can be brought to the form of a confluent Heun equation by setting:
\begin{equation}
 \label{solutionn0}
 f(z) = z^{\frac{|m|}{2}} (1-z)^\ell \mathfrak{f}(z) \; .
\end{equation}
Indeed, if we do this the function $\mathfrak{f}$ has to satisfy the following DE:
\begin{equation}
\label{ConfluentHeun}
 0 = \partial^2_z \mathfrak{f} + \Big(  \frac{\g}{z} + \frac{\d}{z - 1} + \varepsilon  \Big) \partial_z \mathfrak{f} + \frac{\a z - q}{z (z - 1)} \mathfrak{f} \, ,
\end{equation}
with
\begin{equation}
 \g = |m| + 1 \, , \quad \d = 2 \big( \ell + 1 \big) \, , \quad \varepsilon = 0 \, , \quad \a = - \frac{n^2}{16} \, , \quad q = \frac{M^2}{4} - \frac{n^2}{16} - \Bigg(  \frac{|m|}{2} + \ell + 1 \Bigg)^2 + 1 \, ,
\end{equation}
which is the confluent Heun equation.

Eq. 4.7 can also be put in a form of a Schr\"odinger-like problem. To do this we redefine $f(\a)$ as:
\begin{equation}
 f(\a) = \frac{\tilde{\mathfrak{f}}(\a)}{2 \; \cos^{3/2}\a \sin^{1/2}\a} \, .
\end{equation}
Then the function $\tilde{\mathfrak{f}}(\a)$ satisfies the Schr\"odinger equation:
\begin{equation}
 - \partial^2_{\a} \tilde{\mathfrak{f}} + V(\a) \tilde{\mathfrak{f}} = M^2 \tilde{\mathfrak{f}} \, ,
\end{equation}
where the potential $V(\a)$ is:
\begin{equation}
 V(\a) = \frac{4 m^2 - 1}{4 \sin^2\a} + \frac{16 \ell (\ell + 1) + 3}{4 \; \cos^2\a} + \frac{n^2}{4} \cos^2\a - 4 \, .
\end{equation}
Potentials of this type were also considered in the study of marginally deformations of supersymmetric backgrounds \cite{Hernandez:2005xd,Hernandez:2005zx}. To the best of our knowledge, it is still not known how to solve analytically eigenvalue problems with potentials like the one above and thus we will use numerical methods.

\subsubsection{The analytic case $n = 0$}
\label{sec:ATDanalytic}

Let us consider now the case where $n=0$. It is easy to see that the confluent Heun equation \eqref{ConfluentHeun} reduces to a hypergeometric differential equation:
\begin{equation}
\label{HypergeometricDE}
 0 = z (1-z) \partial^2_z \mathfrak{f} + \big[  c - (a + b + 1) z  \big] \partial_z \mathfrak{f} - a b \mathfrak{f} \, ,
\end{equation}
where the constants $a, b, c$ are given in terms of the eigenvalues $\ell,m$ and the mass $M$ through the following relations:
\begin{equation}
 a = 1 + \ell + \frac{|m|}{2} - \sqrt{1 + \frac{M^2}{4}} \; , \quad b = 1 + \ell + \frac{|m|}{2} + \sqrt{1 + \frac{M^2}{4}} \; , \quad c = |m| + 1 \; .
\end{equation}
The hypergeometric equation above admits two linearly independent solutions. Since in our case $c$ is a non-negative integer one of the two solutions is singular at $z = 0$ and thus we will not consider it. Hence the only permissible solution is:
\begin{equation}
 \mathfrak{f}(z) = {}_2 F_1(a,b;c;z) \, .
\end{equation}
However in our case $c - a - b = -2\ell-1 < 0$. For this reason the behavior of the hypergeometric function near $z = 1$ is given by the formula \cite{Olver}:
\begin{equation}
 \lim\limits_{z \rightarrow 1^{-}} \frac{{}_2 F_1(a,b;c;z)}{(1 - z)^{c - a - b}} = \frac{\G(c) \G(a + b - c)}{\G(a) \G(b)} \, .
\end{equation}
Thus the only way to make the solution \eqref{solutionn0} regular at $z = 1$ is to require that $a = - \nu$ with $\nu = 0, 1, 2, \ldots \, .$ From this condition we end up with the following tower of masses and conformal dimensions:
\begin{equation}
\label{ATDanalyticSpec}
 \begin{aligned}
  & M^2 = \Big(  2(\nu + \ell) + |m| \Big) \Big(  2(\nu + \ell + 2) + |m| \Big) \, , \qquad \D =  2(\nu + \ell + 2) + |m| \, ,
  \\[5pt]
  & m \in \mathbb{Z} \, , \qquad \nu, \ell = 0, 1, 2, \ldots \, .
 \end{aligned}
\end{equation}
Taking $\kappa = 2(\nu + \ell) + |m|$ we can write the above formula as:
\begin{equation}
 M^2 = \kappa \big(  \kappa + 4 \big) \, , \qquad \kappa = 0, 1, 2, \ldots \, .
\end{equation}
The last formula matches the result found in \cite{Kim:1985ez} for the excitations of the metric along the $AdS_5$ directions in the case of $AdS_5 \times S^5$. This is expected as the modes with $n = 0$ that we considered here are inert under the T-duality transformation.

\subsubsection{The non-analytic case $n \ne 0$}

As it was already mentioned, we are not aware of any method that allows us to solve eq. \eqref{DEtoSolve} analytically for $n \ne 0$. For this reason we restrict ourselves to find an approximate formula for the masses $M$ which is valid for large enough values of $M$. This can be done using the WKB method of the appendix \ref{WKBmethod}.

In order to be able to apply the WKB method we first have to express eq. \eqref{DEtoSolve} in terms of a suitable variable which we call $r$ and it is related to $z$ as:
\begin{equation}
 r = \frac{z}{1 - z} \, , \qquad r \in [0,+\infty )  \, .
\end{equation}
The next step is to bring eq. \eqref{DEtoSolve} into the form \eqref{WKBDE}. As a result, the functions $p(r), w(r)$ and $q(r)$ of eq. \eqref{WKBDE} are:
\begin{equation}
 p(r) = \frac{r}{1 + r} \, , \qquad w(r) = \frac{1}{4 \, (  1 + r )^3} \, , \qquad q(r) = - \frac{m^2}{4 r (1 + r)^2} - \frac{n^2 + 16 \ell (\ell + 1) (1 + r)^2}{16 (1 + r)^4} \, .
\end{equation}
Expanding these functions in the vicinity of the two end-points we find:
\begin{equation}
 \begin{aligned}
  & p(r) = r + \cO(r^2) \, , \qquad w(r) = \frac{1}{4} + \cO(r) \, ,
  \\[5pt]
  & q(r) = - \frac{m^2}{4 r} + \frac{m^2}{2} - \frac{n^2}{16} - \ell (\ell + 1) + \frac{8 \ell (\ell + 1) + n^2 - 3 m^2}{4} \, r + \cO(r^2)
 \end{aligned}
 \qquad \textrm{at} \quad r \approx 0
\end{equation}
and
\begin{equation}
 \begin{aligned}
  & p(r) = 1 + \cO(r^{-1}) \, , \qquad w(r) = \frac{1}{4 r^3} + \cO(r^{-4}) \, ,
  \\[5pt]
  & q(r) = - \frac{\ell (\ell + 1)}{r^2} + \frac{8 \ell (\ell + 1) - m^2}{4 r^3} + \frac{8 m^2 - n^2 - 48 \ell (\ell + 1)}{16 r^4} + \cO(r^{-5})
 \end{aligned}
 \qquad \textrm{at} \quad r \approx + \infty \, .
\end{equation}
The reason for keeping more terms in the expansions of the function $q(r)$ is to exploit all the different possibilities that one can obtain from its asymptotic behavior. More specifically, one can consider the cases where $(m \ne 0 , \ell \ne 0)$ or $(m \ne 0 , \ell = 0)$ or $(m = 0 , \ell \ne 0)$ or $(m = 0 , \ell = 0)$. It turns out that all of them give the same WKB formula for the masses, which is:
\begin{equation}
\label{MassesWKB}
 M^2 = 4 \nu (\nu + |m| + 2 \ell) \, , \qquad \nu = 1, 2, \ldots \, .
\end{equation}
Notice that the previous formula does not depend on the quantum number $n$. 

As an independent check, we solved eq. \eqref{DEtoSolve} numerically (using the shooting method) for given values of the quantum numbers $m, n$ and $\ell$. As it can be seen from the figure below, the values for the masses that are computed numerically are in a good agreement with those computed by the WKB formula \eqref{MassesWKB} when the mass is large enough.
\begin{figure}[H]
\centering
 \includegraphics[width=0.65\textwidth]{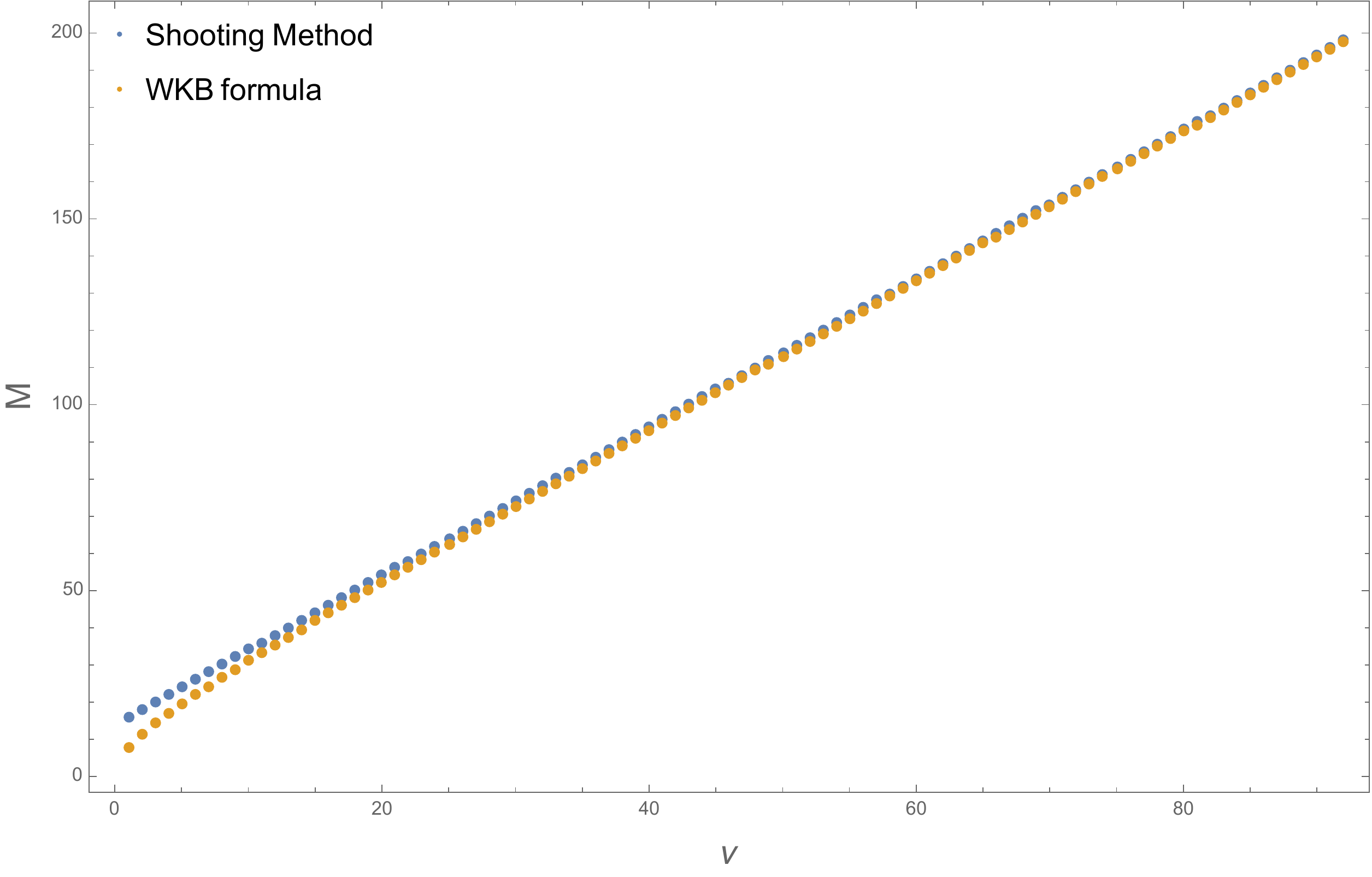}
 \captionsetup{width=0.6\textwidth}
 \caption{\small Comparison of the masses that are computed numerically with the ones computed using the WKB formula. For the computation we fix the quantum numbers as $(n,m,\ell) = (7,4,5)$.}
\end{figure}

In order to illustrate that the masses do not depend on the quantum number $n$ as $M$ is getting higher and higher, we plot the tower of masses for fixed values of $m$ and $\ell$ and different values of $n$. In the figure below, each line corresponds to a different value of $n$. It turns out that the lines tend to merge in the sector of large masses.
\begin{figure}[H]
\centering
 \includegraphics[width=0.65\textwidth]{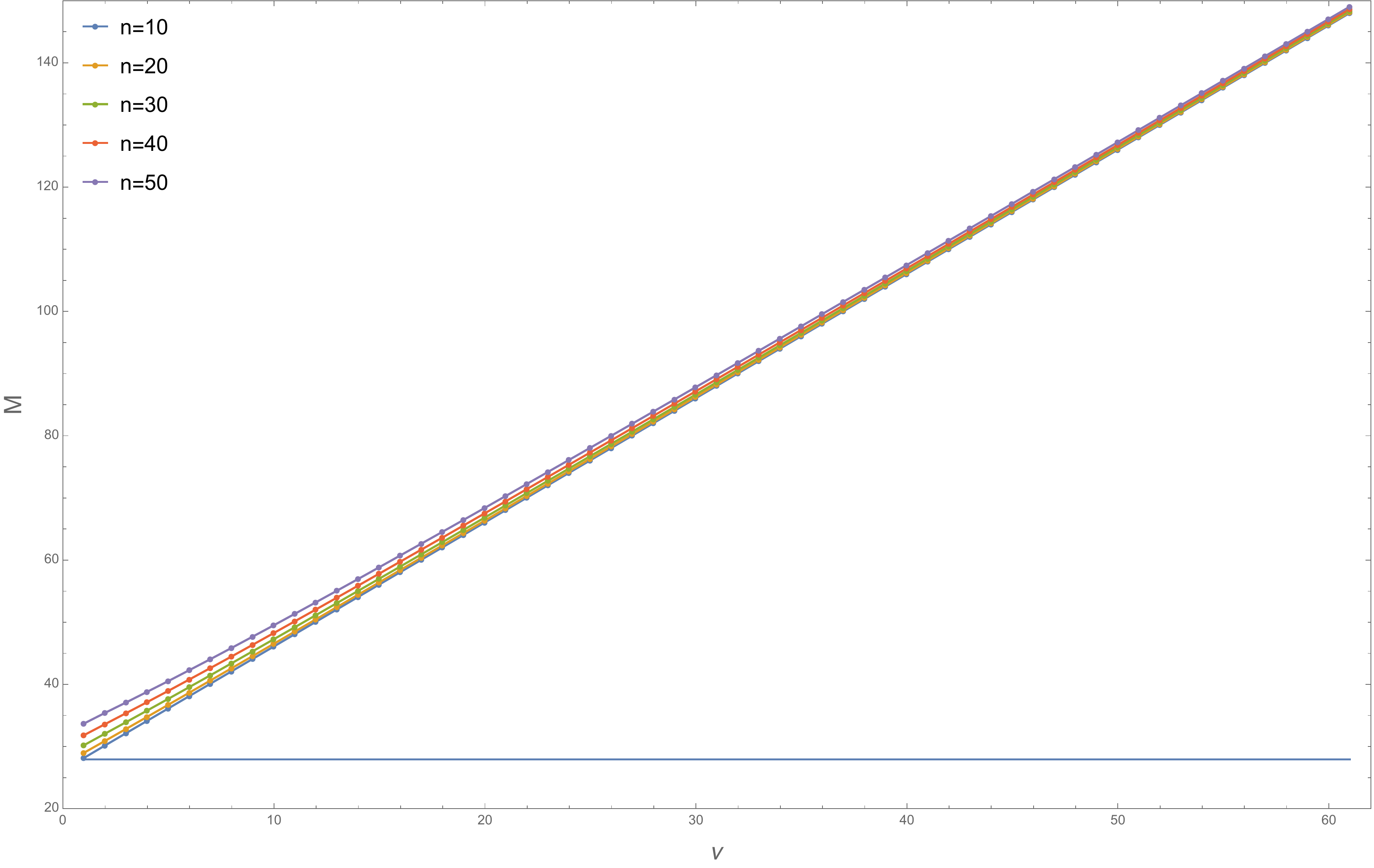}
 \captionsetup{width=0.6\textwidth}
 \caption{\small Mass spectra for fixed $m$ and $\ell$ ($m=10, \ell = 8$) and different values of $n$. The horizontal line represents the lower bound for the masses given in eq. \eqref{Mbound}.}
 \label{figVarn}
\end{figure}

The above feature is not manifest when we fix $n$ and $m$ and vary $\ell$ or when we vary $m$ while keeping $n$ and $\ell$ fixed. This can be seen in the following plots where any possible convergence of the different lines seems to happen much slower compared to the figure \ref{figVarn}.
\begin{figure}[H]
\centering
\captionsetup{width=0.9\textwidth}
\begin{subfigure}{0.49\textwidth}
 \includegraphics[width=\textwidth]{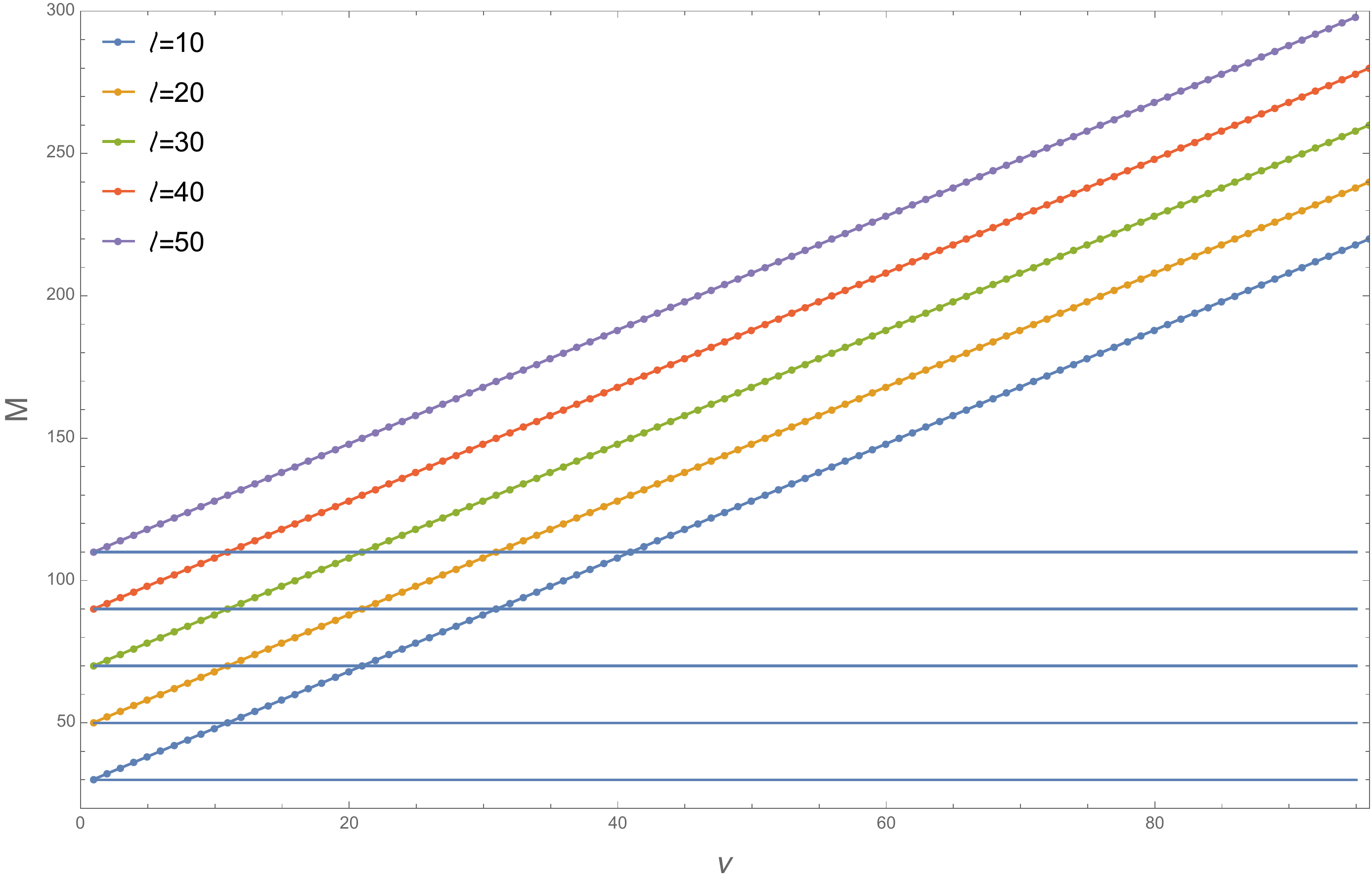}
\end{subfigure}
\,
\begin{subfigure}{0.49\textwidth}
 \includegraphics[width=\textwidth]{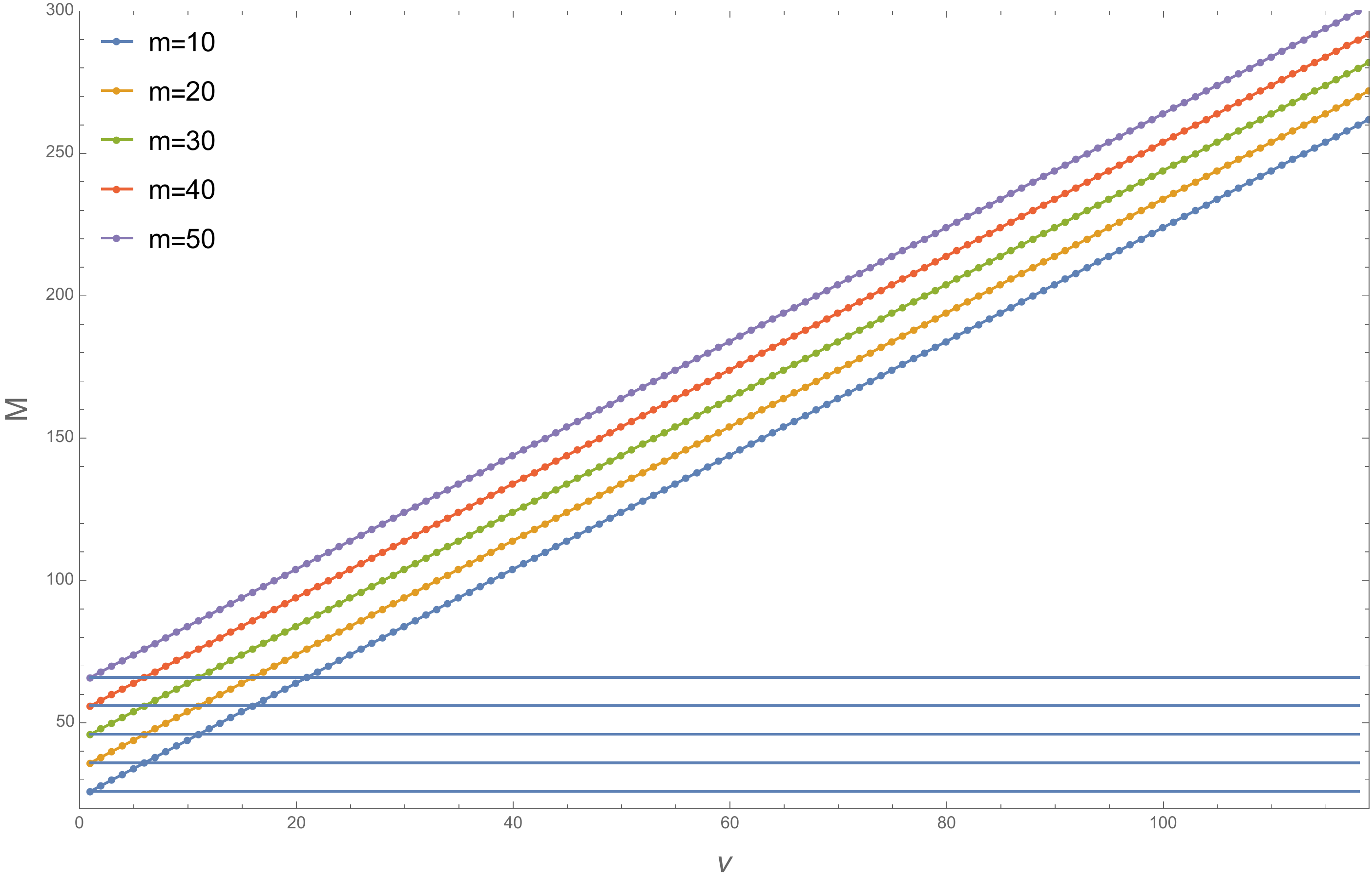}
\end{subfigure}
\caption{\small The figure on the left represents the mass spectra for $n=10, m=8$ and different values of $\ell$. The figure on the right shows the mass towers for $n=4, \ell=7$ and different values of $m$. The horizontal lines represent the lower bounds for the masses given in eq. \eqref{Mbound}.}
 \label{figVarlm}
\end{figure}

\subsubsection{Bound on the spectrum}

We will now recast eq. \eqref{ConfluentHeun} in the Sturm-Liouville fashion:
\begin{equation}
\label{SL}
 0 = S \mathfrak{f} + \l W(z) \mathfrak{f} \, , \qquad S := \frac{d}{dz} \Bigg(  P(z) \frac{d}{dz} \Bigg) + Q(z) \, .
\end{equation}
This is done for:
\begin{equation}
 P(z) = z^{|m|+1} (1 - z)^{2 \ell + 2} \, , \qquad Q(z) = - \frac{n^2}{16} \, z^{|m|} (1 - z)^{2 \ell + 2} \, , \qquad W(z) = z^{|m|} (1 - z)^{2 \ell + 1}
\end{equation}
and
\begin{equation}
 \l = \frac{1}{4} \Big[ M^2 - \big( 2 \ell + |m|  \big) \big(  2 \ell + |m| + 4 \big)  \Big] \, .
\end{equation}

Let us also introduce the following inner product with respect to the weight function $W(z)$:
\begin{equation}
 \big< \mathfrak{f}_1,\mathfrak{f}_2 \big>_W := \int\limits_{0}^{1} dz \, W(z) \, \mathfrak{f}_1(z) \, \mathfrak{f}_2(z) \, .
\end{equation}
We will impose boundary conditions such that two eigenfunctions of different eigenvalues are orthogonal. Then we have:
\begin{equation}
 \begin{aligned}
  & 0 = S \mathfrak{f}_1 + \l_1 W(z) \mathfrak{f}_1 \, ,
  \\[5pt]
  & 0 = S \mathfrak{f}_2 + \l_2 W(z) \mathfrak{f}_2 \, .
 \end{aligned}
\end{equation}
If we multiply the first by $\mathfrak{f}_2$ and the second by $\mathfrak{f}_1$ and then subtract them we find:
\begin{equation}
 0 = \mathfrak{f}_2 \frac{d}{dz} \Bigg(  P(z) \frac{d \mathfrak{f}_1}{dz} \Bigg) - \mathfrak{f}_1 \frac{d}{dz} \Bigg(  P(z) \frac{d \mathfrak{f}_2}{dz} \Bigg) + \big(  \l_1 - \l_2 \big) W(z) \, \mathfrak{f}_1 \, \mathfrak{f}_2 \, .
\end{equation}
Integrating the last from $0$ to $1$ we get:
\begin{equation}
 \big(  \l_1 - \l_2 \big) \big< \mathfrak{f}_1,\mathfrak{f}_2 \big>_W = P(z) \, \Bigg(  \mathfrak{f}_1 \, \frac{d \mathfrak{f}_2}{dz} - \mathfrak{f}_2 \, \frac{d \mathfrak{f}_1}{dz} \Bigg) \Bigg|^{z = 1}_{z = 0} \, .
\end{equation}
Hence we impose:
\begin{equation}
 P \, \mathfrak{f} \, \frac{d \mathfrak{f}}{dz} \Bigg|_{z=0} = P \, \mathfrak{f} \, \frac{d \mathfrak{f}}{dz} \Bigg|_{z=1} = 0 \, ,
\end{equation}
which in our case is satisfied as long as $\mathfrak{f}$ and $d\mathfrak{f}/dz$ are finite at the endpoints of the interval $[0,1]$, or if they diverge they do it slow enough.

We can now derive a lower bound for the mass spectrum in the following way. From the Sturm-Liouville equation \eqref{SL} we have:
\begin{equation}
 \l \, \big< \mathfrak{f},\mathfrak{f} \big>_W = - \int\limits_{0}^{1} dz \, \mathfrak{f} \, S \mathfrak{f} =  \int\limits_{0}^{1} dz \, P \, \Bigg(  \frac{d \mathfrak{f}}{dz} \Bigg)^2 - \int\limits_{0}^{1} dz \, Q \, \mathfrak{f}^2 \ge 0 \, ,
\end{equation}
which implies that $\l \ge 0$ or:
\begin{equation}
\label{Mbound}
 M^2 \ge \big( 2 \ell + |m|  \big) \big(  2 \ell + |m| + 4 \big) \, .
\end{equation}
Notice that the bound does not depend on the number $n$.

 Obviously the spectrum we found in eq. \eqref{ATDanalyticSpec} satisfies the above bound and it saturates it for $\nu = 0$. From the numerical analysis we get strong evidence that the bound is not violated for non-zero values of $n$, as it can be seen in the figures \ref{figVarn} and \ref{figVarlm}.

\subsection{The NATD case}

Though in the non-Abelian T-dual case things seem to be similar to the Abelian example, one has to be careful with the range of the coordinate $\rho$ as it runs in the semi-infinite interval $[0,+\infty)$. 
The unboundedness of the $\rho$ coordinate makes the NATD solution a non-regular solution of the Laplace equation (\ref{condV}) since it does not satisfy the last of the boundary conditions in eq. (\ref{condV2}). 
We will see in this section that our analysis will set a bound to this coordinate. 

The separation of variables scheme for the operator $\cL^{NATD}$ is:
\begin{equation}
 \begin{aligned}
  & Y(y) = \sum\limits_{m,\ell,k} \int\limits_{0}^{\infty} dn \, f_{m,n,\ell}(\a) e^{i m \b} j_\ell(n \rho) \cY^k_{\ell}(\chi, \xi) \, ,
  \\[5pt] 
  & m \in \mathbb{Z} \, , \qquad n \in \mathbb{R}_{\ge 0} \, , \quad \ell = 0, 1, 2, \ldots \, , \quad k = - \ell , - \ell + 1 , \ldots , \ell \, ,
 \end{aligned}
\end{equation}
where $j_\ell(n \rho)$ are the spherical Bessel functions that are regular at the origin. This scheme leads to the differential equation \eqref{EVATD} where now the index $n$ is continuous. As a result one expects a non-discrete spectrum of masses with respect to the index $n$. Notice however that the space of admissible solutions of the eigenmode equation must involve normalizable eigenfunctions. 
That is to say that the measure $\cC_{m, \ell, k, n}$ in the effective action for the 5d graviton must be finite. The effective action results from the second variation of the type-IIA supergravity action and in our case it is
\begin{equation}
 \begin{aligned}
  & \d^2 S_{IIA} = \sum\limits_{m, \ell, k} \int\limits_0^{\infty} dn \, \cC_{m, \ell, k, n} \int d^5 x \sqrt{- \tilde{g}_{AdS_5}} \big(  h^{[tt]}_{m, \ell, k, n}  \big)^{\m\n} \Big[  \square_{AdS_5} + \big(  2 - M^2  \big)  \Big] \big(  h^{[tt]}_{m, \ell, k, n}  \big)_{\m\n} \, ,
  \\[5pt]
  &\qquad \qquad \qquad \qquad  \cC_{m, \ell, k, n} = \frac{1}{\kappa^2_{10}} \int\limits_{\cM_5} d^5 y \sqrt{\tilde{g}_{\cM_5}} \, e^{8A} \, \big|  Y_{m, \ell, k, n} \big|^2 \, .
 \end{aligned}
\end{equation}
It turns out that $\cC_{m, \ell, k, n}$ involves an integral of the form $\int\limits_0^{\infty} d\r \r^2 \big( j_{\ell} (n \r) \big)^2$ which diverges unless the integration is done on a finite interval.
Finite perturbations then instruct us to impose a bound for $\rho$ which corresponds to a hard cut-off in the geometry. 
Let us work for the moment with the solution with finite $\rho$ and leave some comments to the end of the section.

Since now $\rho$ is of finite range one has to make a choice for the boundary conditions of the function $Y(y)$ at $\rho = \rho_*$. 
 Let us assume for example that $Y|_{\rho = \rho_*} = 0$. This implies that $n$ can take only those values where:
\begin{equation}
 n_{\ell s} = \frac{\rho_{\ell s}}{\rho_*} \, , \qquad s = 1, 2, \ldots \, ,
\end{equation}
with $\rho_{\ell s}$ being the roots of $j_{\ell}(\rho)$. The separation of variables scheme now reads:
\begin{equation}
 \begin{aligned}
  & Y(y) = \sum\limits_{m,\ell,k, s} f_{m,\ell, s}(\a) e^{i m \b} j_\ell(n_{\ell s} \rho) \cY^k_{\ell}(\chi, \xi) \, , 
  \\[5pt]
  & m \in \mathbb{Z} \, , \quad s = 1, 2, \ldots \, , \quad \ell = 0, 1, 2, \ldots \, , \quad k = - \ell , - \ell + 1 , \ldots , \ell \, .
 \end{aligned}
\end{equation}

Again, the functions $f_{m,n,\ell}(\a)$ must satisfy the DE \eqref{EVATD}, but now the parameter $n$ has to be replaced by $n_{\ell s}$. Consequently, when $n_{\ell s} = 0$, it is possible to find analytic solutions for \eqref{EVATD}. Since we are dealing with the same DE, one expects to find the same mass spectrum as in eq. \eqref{ATDanalyticSpec}. In the case where $n_{\ell s} \ne 0$ the DE \eqref{EVATD} can only be solved numerically. Since the only difference with the previous examples is the fact that now we have to deal with the values of $n_{\ell s}$ instead of the integer or positive real values of $n$, we do not attempt such an analysis. For large quantum numbers one can perform a WKB analysis which should give as a result the same behavior for the masses as in eq. \eqref{MassesWKB}.
Moreover, the mass bound for the aforementioned two examples turns out to be the same as in eq. \eqref{Mbound}. This is due to the fact that one has to deal with the same DE \eqref{DEtoSolve} as in the ATD case.

Let us finally point out the problems regarding the finiteness of the $\rho$ coordinate in the NATD solution. In principle, there is no reason to end the spacetime at some finite point $\rho_{*}$ consistently 
without adding extra sources to the solution. Let us briefly explain this. As we pointed out in the introduction, generic D-brane configurations describing MG backgrounds involve an arrangement of  D4-NS5-D6 branes. 
In this vein the NATD solution corresponds to an infinite set of NS5 branes located at some finite positions along $\rho$ with $pN_6$ D4-branes stretched between the $p^{\textrm{th}}$-$(p+1)^{\textrm{th}}$ NS5 branes. 
There are not D6-branes in the solution. This can be seen from the fact that the potential function describing the NATD solution eq. (\ref{VST}) gives a charge density which is an ever increasing (non piecewise)
 linear function\footnote{we have perfomed a rescaling in the metric such that $\lambda\rightarrow N_6\lambda$ \cite{Gaiotto:2009gz}.} 
$\lambda(\eta)=N_6\eta$ with zero change in slope, $a_{i}-a_{i-1}=0$, signalling the absence of D6 branes (see discussion around eq. (\ref{condV}). It is then evident that a decreasing change in slope such that the $\rho$ 
coordinate is of finite size can be ascertained by adding D6 branes to the solution. The solution of the Laplace equation $V(\eta,\sigma)$ is then modified. 
A generic solution satisfying the appropriate boundary conditions can be written as \cite{Aharony:2012tz}
 \begin{equation}
 \begin{split}\label{genericMG}
V(\sigma,\eta)=&-\sum_{n=1}^{\infty}\frac{c_n}{\omega_n}K_{0}(\omega_n\sigma)\sin (\omega_n\eta),\qquad \omega_n=\frac{n\pi}{\eta_{*}}, \quad 0\leq \eta\leq \eta_{*},\\
&\qquad \lambda(\eta)=\dot{V}(0,\eta)=\sum_{n=1}^{\infty} c_{n}\sin\omega_{n}\eta, 
\end{split}
\end{equation}
where $c_n$ are the Fourier coefficients associated to $\lambda(\eta)$.  Eventhough our analysis in this section was based on the NATD solution one can prove that the bound in the mass found in eq. (\ref{Mbound}), 
and correspondingly the dimension for the operators, is universal for the MG class of solutions described by eq. (\ref{genericMG}) in agreement with the results in \cite{Chen:2019ydk}. In addition to this, it can be shown that
 the solution in eq. \ref{genericMG} -using Poisson resumation formula- close to $\sigma,\eta\sim0$ behaves as the NATD solution (see e.g.  \cite{Nunez:2019gbg} for details). This suggest that the NATD solution gives a good description of the physics as long as we are in this parameter region, far away from the end of the space determined by the finite range of $\eta$ (or $\rho$ in our notation). Interestingly, the absence of D6 branes in the NATD solution is the key feature for
 integrability \cite{Nunez:2018qcj}, contrary to the generic solutions of the MG class. 
We will propose in the next section operators belonging to this integrable subsector, that is to say when fundamental fields are not considered.

\section{Some comments about the field theory interpretation}
In this section we close with some comments related to the field theory interpretation of our analysis. 

\begin{figure}[H]
\centering
 \includegraphics[width=0.8\textwidth]{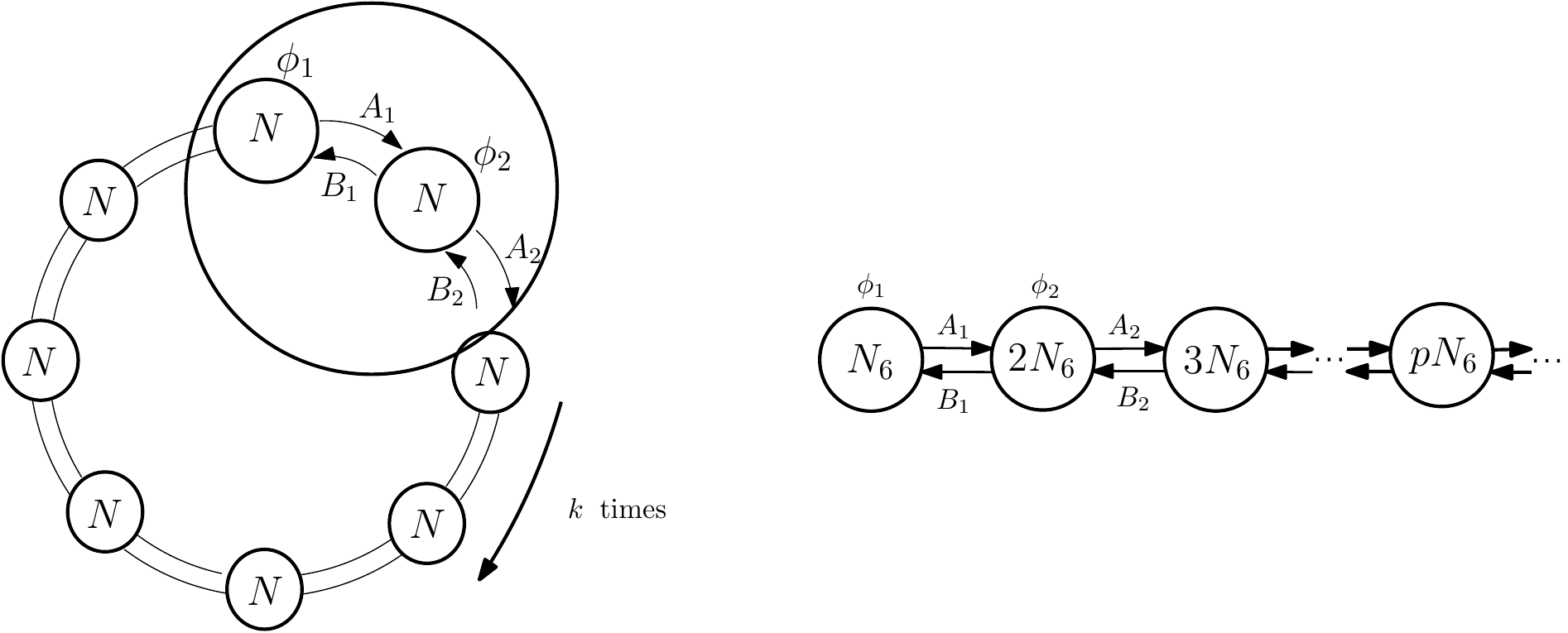}
 \captionsetup{width=0.6\textwidth}
 \caption{\small Dual quivers associated to the ATD (left) and NATD (right) solutions. We use $\mathcal{N}=1$ language to represent $\mathcal{N}=2$ hypermultiplets with two arrows.}
 \label{lastpic}
\end{figure}

Generic SCFTs dual to MG class of solutions are described by linear quivers with product gauge groups $\prod_{i=1}^{n} SU(N_i)$ and 4d $\mathcal{N}=2$ supersymmetry 
with $SU(2)_R\times U(1)_r$ R-charge. The field content of the theory involves a vector multiplet for each gauge node as well as hypermultiplets transforming in the
bifundamental representation of consecutive gauge groups. In addition we have fundamental fields attached to particular gauge nodes.

We studied the spin-2 spectrum associated to two interesting examples dual to the above SCFTs. Following the rules spelled out in \cite{Gaiotto:2009gz} the quivers encoding the information of the field theory duals 
to the ATD and NATD solutions were worked out in \cite{Lozano:2016kum}. The field theory dual to the ATD is described by a circular quiver and describes the $\mathbb{Z}_k$ orbifold projection of $\mathcal{N}=4$ SYM.
 The dual quiver for the NATD solution is given by a linear quiver of infinite size (see figure \ref{lastpic}). 
In the following, we will mostly concentrate the discussion on the NATD case.

As pointed out, the dual field theory to the NATD solution corresponds to an infinitely long linear quiver with gauge group $\prod_{p=1}^{\infty} SU(p N_6)$. The infiniteness of the linear quiver represents a problem
whenever we compute certain physical observables \cite{Lozano:2016kum}. The results of the preceding section showed that in order for the spin-2 spectrum to have a well-defined Hilbert space for the eigenfunctions 
of the wave operator it is necessary to restrict the infinite range of the $\rho$ coordinate to a finite one. The boundedness of the $\rho$ coordinate in the geometry implies, for the dual field theory, the finiteness of the linear quiver. 
A consistent way to obtain a finite quiver is ascertained by adding extra fundamental fields attached, for instance, to the last gauge node the number of which such that we preserve conformal symmetry. 
This is yet another way to see that the field theory dual to the NATD needs to be {\it completed} and is along the lines of the CFT completions studied in \cite{Lozano:2016kum}.

In the previous section we showed that both ATD and NATD solutions (for $\nu=0$) have spin-2 operators with dimension $\Delta=4+2l+\vert m\vert$ where $\ell$ and $m$ are identified with the $SU(2)$ and $U(1)$ spins respectively.
 The structure of the multiplets containing the spin-2 states with the corresponding dimension dual to the MG class of solutions were identified in \cite{Chen:2019ydk}.
Using the notation $[j,\overline{j}]_{\Delta}^{(R,r)}$ where $j,\overline{j}$ label the representation of the conformal algebra, the two-sided chiral multiplets and the representative primaries are
\begin{equation}
A_2A_2:\quad [0,0]^{(2\ell,0)}_{\Delta=2+2\ell},\quad A_2\overline{L}:\quad [0,0]^{(2\ell,-2m)}_{\Delta=2+2\ell+2m}, \quad L\overline{A_2} :\quad [0,0]^{(2\ell,2m)}_{\Delta=2+2\ell+n}.
\end{equation}
We can then use the field content of the theories we studied here in order to construct operator candidates dual to the spin-2 states belonging to the above multiplets. 
Intuitively they may involve a gauge invariant combination of fields times the stress energy momentum tensor which is dual to the massless spin-2 state and has dimension\footnote{following the conventions of \cite{Cordova:2016emh} that we will use from now on} $\Delta=4$.

Let us start with the ATD case. As discussed above the dual quiver is circular containing $k$ gauge groups and  bifundamental mater. We then have $k$ different colour indices associated to each scalar in each of the vector multiplets, $\phi^{a_{i}}_{b_{i}}$, as well as bifundamentals with $k$ indices 
 of different kind between adjacent gauge nodes, $A^{a_{i}}_{b_{i+1}}$.  The above fields have dimension 1 and R-charges $\phi: (0,2), \, A: (1,0)$ under $SU(2)_R\times U(1)_r$
 respectively. We can then consider operators of the schematic form
 \begin{equation}\label{opnatd}
 \mathcal{O}_{\mu\nu}=\textrm{Tr}(\phi_{1}^{n_1} A_{1}^{I_1} \phi_{2}^{n_2} A_{2}^{I_2}\ldots \phi_{k}^{n_k} A_{k}^{I_{k}})T_{\mu\nu},\quad  \sum_{i=1}^{k}n_{i}=m,\quad  \sum_{i=1}^{k}I_{i}=2\ell,
 \end{equation}
with R-charge $(2l,2m)$  where $I$ is an $SU(2)_R$ index. This operator may be thought of as generalizations of the BMN operators studied in \cite{Itsios:2017nou}. In the NATD solution this kind of operators are
 not well-defined since they will have a very long dimension -infinite- and therefore they will be so heavy to be excited. We do not expect them to be  part of the 4D $\mathcal{N}=2$ dynamics.
 This is in fact due to the infiniteness of the quiver describing the dual SCFT. As we discussed before in order for the quiver to be finite we need to add extra fundamental fields. 
 
 In the absence of fundamental fields, there are still well-defined operators that we can study. As we pointed out at the end of the last section, the NATD solution is the universal behaviour of any solution
 belonging to the MG class close to $\sigma,\eta\sim 0$. One may then expect to define operators in the regime where we can ignore the presence of flavours.
  For instance, we can form mesons with the bifundamentals in the hypers $\mathcal{M}^{a}_{b}=A^{a}_{\alpha} B^{\alpha}_{b}$. A spin-2 operator then may have the form
 of a {\it dimeric}-like operator  \cite{Gadde:2010zi}
 \begin{equation}\label{opnatd2}
 \mathcal{O}_{\mu\nu}=\textrm{Tr}(\phi^m\mathcal{M}^{\ell})T_{\mu\nu},
 \end{equation}
with R-charge $(2\ell,2m)$. We argue this operator belongs to the integrable subsector of generic SCFTs dual to MG class of solutions according with the results of \cite{Nunez:2018qcj}.

To conclude this section let us mention that the above analysis goes through also for $\nu \neq 0$. In this case we can add $SU(2)_R$ or $U(1)_r$ singlets inside the trace of the operators
(\ref{opnatd}) and (\ref{opnatd2}) to achieve the $\nu$ dependence in the dimension.

\section{Discussion}
In this paper we have written down the whole set of linearized equations of motion for fluctuations of warped geometries in type-IIA supergravity with $AdS_5$ factor. In particular we studied the
spin-2 excitations of the Gaiotto-Maldacena class of geometries. We gave a generic expression for the wave operator describing these fluctuations in terms of the solution to the axisymmetric Laplace equation
characterizing these geometries. We studied two interesting examples: The Abelian (Hopf) T-dual and the Non-Abelian T-dual geometries of the maximally supersymmetric solution $AdS_5\times S^5$. 
The wave operator for these solutions turned out to be the same when the ``field space'' coordinate $\rho$ of the NATD solution is large. 
We were able to find an analytic solution for the spectrum only when the quantum number $n$ vanishes. For the rest of the spectrum we resort on numerical methods.
 For large masses we showed that our results are in perfect agreement with WKB expectations.
 Since in the ATD case the ``field space'' coordinate is compact, we obtained a discrete spectrum of masses. 
 A bound for these masses was found. However, in the NATD case, the ``field space'' coordinate is unbounded, which originates a continuous spectrum of masses.
 By imposing a hard cut-off in the geometry that bounds the value of this coordinate a discrete spectrum of masses emerged. We conclude our work with a field theory interpretation of the spin-2 excitations in the two examples of our interest by proposing dual operators belonging to the multiplets of the SCFTs dual to generic MG class of solutions identified in \cite{Chen:2019ydk}.

An immediate application of the equations for the fluctuations written in the appendices
is to study the full spectrum of excitations for any solution in type-IIA supergravity. In particular, the marginally 
deformed Gaiotto-Maldacena solutions studied in \cite{Nunez:2019gbg}. These geometries are also characterized by the 
function $V(\sigma,\eta)$ that solves the Laplace equation (\ref{condV}). It is then possible to write down generic 
formulas in terms of this function to calculate the full spectrum. We plan to report these results in a future collaboration.

\section*{Acknowledgements}

We are very grateful to H. Nastase, C. N\'u\~nez, A. Ramallo and K. Sfetsos for useful discussions. The work of G.I. is supported by FAPESP grant 2016/08972-0 and 2014/18634-9. J. M. P. is funded by the Spanish grant FPA2017-84436-P, by Xunta de Galicia (GRC2013-024), by FEDER and by the Maria de Maetzu Unit of Excellence MDM-2016-0692, and supported by the Spanish FPU fellowship FPU14/06300.
The work of S. Z. is supported by the National Natural Science Foundation of China (NSFC) grant 11874259.


\appendix

\section{Useful formulas}

In this appendix we list some of the most useful formulas and identities that are necessary to derive the linearized equations of motion for the fluctuations of the supergravity fields. 

\subsection{List of formulas in the Riemannian geometry}

The Christoffel symbols and the covariant derivatives acting on tensors are given by:
\begin{eqnarray}
 && \G^P_{MN} = \frac{1}{2} \, g^{P \Sigma} \, \big(   \partial_M g_{\Sigma N} + \partial_N g_{\Sigma M} - \partial_{\Sigma} g_{M N}  \big) \, ,
 \\[5pt]
 && \nabla_M T^{A_1 \ldots A_n}_{\qquad\;\;\;\; B_1 \ldots B_m} = \partial_M T^{A_1 \ldots A_n}_{\qquad\;\;\;\; B_1 \ldots B_m} + \sum\limits_{k = 1}^{n} \G^{A_k}_{M P} \, T^{A_1 \ldots P \ldots A_n}_{\qquad\qquad B_1 \ldots B_m}
 \\[5pt]
 && \nonumber \qquad\qquad\qquad\qquad\qquad\qquad\qquad\qquad\; 
 - \sum\limits_{k = 1}^{m} \G^{P}_{M B_k} T^{A_1 \ldots A_n}_{\qquad\;\;\;\; B_1 \ldots P \ldots B_m} \, .
\end{eqnarray}
Notice that when the upper index of the Christoffel symbols is contracted with one of its lower indices then:
\begin{equation}
 \G^{M}_{MN} := \frac{1}{2} \, g^{M \Sigma} \, \big(   \partial_M g_{\Sigma N} + \partial_N g_{\Sigma M} - \partial_{\Sigma} g_{M N}  \big) = \frac{1}{2} \, g^{M \Sigma} \, \partial_N g_{M \Sigma} = \partial_N \ln \sqrt{g} \, ,
\end{equation}
where $g := \det \big(  g_{MN} \big)$.

The Riemann tensor is defined through the Christoffel symbols as:
\begin{equation}
 R^{\Lambda}_{\;\;\; M \Sigma N} := \partial_{\Sigma} \G^{\Lambda}_{MN} - \partial_{N} \G^{\Lambda}_{M \Sigma} + \G^{K}_{MN} \G^{\Lambda}_{K \Sigma} - \G^{K}_{M \Sigma} \G^{\Lambda}_{KN} \, .
\end{equation}
In this form it is clear that the Riemann tensor is antisymmetric under $\Sigma \leftrightarrow N$. The Ricci tensor is defined by contracting the first and third indices of the Riemann tensor, i.e.
\begin{equation}
 R_{MN} := R^{\Sigma}_{\;\;\; M \Sigma N} \, .
\end{equation}
Obviously the Ricci tensor is symmetric under the exchange of its indices. Finally the Ricci scalar is given by the contraction of the Ricci tensor with the metric:
\begin{equation}
 R := g^{MN} R_{MN} \, .
\end{equation}
Another useful object is the commutator of two covariant derivatives acting on a tensor. This can be written in terms of the Riemann tensor as:
\begin{equation}
\label{ComCD}
 \big[   \nabla_M, \nabla_N   \big] T^{A_1 \ldots A_n}_{\qquad\;\;\;\; B_1 \ldots B_m} = \sum\limits_{k = 1}^{n} R^{A_k}_{\;\;\;\; \Sigma MN} \, T^{A_1 \ldots \Sigma \ldots A_n}_{\qquad\qquad B_1 \ldots B_m} - \sum\limits_{k = 1}^{m} R^{\Sigma}_{\;\;\; B_k MN} \, T^{A_1 \ldots A_n}_{\qquad\;\;\;\; B_1 \ldots \Sigma \ldots B_m} \, .
\end{equation}


\subsection{Metric variations}

We consider variations around the background metric $\bar{g}_{MN}$ of the form:
\begin{equation}
 g_{MN} = \bar{g}_{MN} + \d g_{MN} \, , \qquad g^{MN} = \bar{g}^{MN} + \d g^{M N} \, ,
\end{equation}
where:
\begin{equation}
 \d g^{M N} = - \bar{g}^{M P} \, \d g_{P \Sigma} \, \bar{g}^{\Sigma N} \, .
\end{equation}

We will also take all the geometric quantities (such as the Christoffel symbols, the covariant derivatives, the Riemann and Ricci tensors and also the Ricci scalar) to be constructed with the background metric $\bar{g}_{MN}$. As a consequence:
\begin{equation}
 \nabla_{P} \bar{g}_{MN} = \nabla_{P} \bar{g}^{MN} = 0 \, .
\end{equation}

The variation of the Christoffel symbols reads:
\begin{equation}
\label{VarChr}
 \begin{aligned}
  & \d \G^{P}_{MN} = \frac{1}{2} \, \bar{g}^{P \Sigma} \, \big(   \nabla_M \d g_{\Sigma N} + \nabla_N \d g_{\Sigma M} - \nabla_{\Sigma} \d g_{M N}\big) \, ,
  \\[5pt]
  & \d \G^{M}_{MN} = \frac{1}{2} \, \bar{g}^{M \Sigma} \,  \nabla_N \d g_{\Sigma M} \, .
  \end{aligned}
\end{equation}

Using the above we can compute the variation of the Riemann tensor which simply becomes:
\begin{equation}
 \d R^{\Lambda}_{\;\;\; M \Sigma N} =  \nabla_{\Sigma} \d \G^{\Lambda}_{MN} - \nabla_{N} \d \G^{\Lambda}_{M \Sigma} \, .
\end{equation}
From this we find that the variation of the Ricci tensor is:
\begin{equation}
 \begin{aligned}
 \d R_{MN} & =  \nabla_{P} \d \G^{P}_{MN} - \nabla_{N} \d \G^{P}_{M P}
 \\[5pt]
 & = \frac{1}{2} \, \big(   \nabla^{\Sigma} \nabla_M \d g_{\Sigma N} + \nabla^{\Sigma} \nabla_N \d g_{\Sigma M} - \nabla^2 \d g_{M N} - \bar{g}^{P \Sigma} \,  \nabla_N \nabla_M \d g_{\Sigma P} \big) \, .
 \end{aligned}
\end{equation}
Finally, for the variation of the Ricci scalar we have:
\begin{equation}
 \begin{aligned}
 \d R & = R_{MN} \, \d g^{MN} + \nabla^{M} \nabla^N \d g_{M N} - \bar{g}^{MN} \, \nabla^2 \d g_{MN}
 \\[5pt]
 & = - R^{MN} \, \d g_{MN} + \nabla^{M} \nabla^N \d g_{M N} - \bar{g}^{MN} \, \nabla^2 \d g_{MN} \, .
 \end{aligned}
\end{equation}


\subsection{Conformal rescalings}

Let us now consider the conformal rescaling of the background metric to be:
\begin{equation}
\label{ConfResc}
 \bar{g}_{MN} = e^{2A} \tilde{g}_{MN} \, .
\end{equation}
Using this, the Christoffel symbols constructed with the background metric $\bar{g}_{MN}$ are given by:
\begin{equation}
\label{ConfChris}
   \G^{P}_{MN} = \tilde{\G}^{P}_{MN} + T^{P}_{MN} \, ,
\end{equation}
where $\tilde{\G}^{P}_{MN}$ are the Christoffel symbols that are constructed with the metric $\tilde{g}_{MN}$ and we defined
\footnote{
We use the fact that for any scalar $f$ it is $\partial_M f = \tilde{\partial}_M f = \nabla_M f = \tilde{\nabla}_M f$.
}
\begin{equation}
 T^{P}_{MN} := \d^P_N \, \tilde{\nabla}_M A + \d^P_M \, \tilde{\nabla}_N A - \tilde{g}_{M N} \, \tilde{\nabla}^{P} A \, .
\end{equation}
Notice that $T^{P}_{MN}$ is symmetric in its lower indices, i.e. $T^{P}_{MN} = T^{P}_{NM}$. Moreover here we raise/lower the indices using $\tilde{g}_{MN}$.

Using the above we can write the Riemann tensor as:
\begin{equation}
 \begin{aligned}
  R^{\Lambda}_{\;\;\; M \Sigma N} & = \tilde{R}^{\Lambda}_{\;\;\; M \Sigma N} + \Big[   \d^{\Lambda}_N \, \tilde{\nabla}_{\Sigma} \tilde{\nabla}_M A - \tilde{g}_{MN} \, \tilde{\nabla}_{\Sigma}\tilde{\nabla}^{\Lambda} A + \d^{\Lambda}_{\Sigma} \, \tilde{\nabla}_N A \, \tilde{\nabla}_M A
  \\[5pt]
  & - \d^{\Lambda}_{\Sigma} \, \tilde{g}_{MN} \, \big(  \tilde{\nabla} A \big)^2 + \tilde{g}_{MN} \, \tilde{\nabla}_{\Sigma} A \, \tilde{\nabla}^{\Lambda} A - \big(  \Sigma \leftrightarrow N  \big)   \Big] \, .
 \end{aligned}
\end{equation}
The Ricci tensor is found after contracting $\Lambda$ with $\Sigma$ in the expression above giving:
\begin{equation}
 \begin{aligned}
  R_{MN} & = \tilde{R}_{MN} + \big(  2 - \cD  \big) \, \tilde{\nabla}_M \tilde{\nabla}_N A - \tilde{g}_{MN} \, \tilde{\nabla}^2A 
  \\[5pt]
  & + \big(  \cD - 2 \big) \, \tilde{\nabla}_M A \, \tilde{\nabla}_N A + \big( 2 - \cD  \big) \,\tilde{g}_{MN} \, \big(\tilde{\nabla} A)^2 \, ,
 \end{aligned}
\end{equation}
where $\cD$ is the dimension of the spacetime. From this we can obtain the Ricci scalar:
\begin{equation}
\label{RicciConf}
 R = e^{-2A} \, \Big[  \tilde{R} - 2 \, \big(  \cD - 1  \big) \, \tilde{\nabla}^2A - \big(  \cD - 1 \big) \, \big(  \cD - 2 \big) \, \big(  \tilde{\nabla} A  \big)^2  \Big] \, .
\end{equation}

Another useful quantity is the variation of the Christoffel symbols in terms of the metric $\tilde{g}_{MN}$. From the first eq. in \eqref{VarChr} this is found to be:
\begin{equation}
\label{VarChrConf}
 \begin{aligned}
  \d \G^{P}_{MN} = \frac{1}{2} \, \tilde{g}^{P \Sigma} \, \big(   \tilde{\nabla}_M h_{\Sigma N} + \tilde{\nabla}_N h_{\Sigma M} - \tilde{\nabla}_{\Sigma} h_{M N} - 2 \, \tilde{g}_{MN} \, h_{\Sigma K} \, \tilde{\nabla}^K A - 2 \, h_{M N} \, \tilde{\nabla}_{\Sigma} A \big) \, ,
 \end{aligned}
\end{equation}
where for convenience we set:
\begin{equation}
 \d g_{MN} = e^{2A} h_{MN} \qquad \Rightarrow \qquad \d g^{MN} = - e^{2A} h^{MN} \, .
\end{equation}
Notice that the indices are raised with $\bar{g}^{MN}$. The expression \eqref{VarChrConf} is useful when we want to express variations of covariant derivatives acting on tensors in terms of the metric $\tilde{g}_{MN}$.


\section{The type-IIA supergravity equations of motion and their fluctuations}

In this appendix we review the equations of motion of the type-IIA supergravity and present general formulas for their fluctuations.

\subsection{The equations of motion of the type-IIA supergravity}

Let us start by writing the equations of motion for the type-IIA supergravity fields \cite{Passias:2018swc}:
\begin{align}
\nonumber
\label{EinsteinEq}
 0 & = R_{MN} - \frac{1}{2} \partial_M \Phi \, \partial_N \Phi - \frac{e^{3\Phi/2}}{2} \Big(  F_{MP} F^{\;\; P}_N - \frac{1}{16} g_{MN} F^2_2  \Big)
 \\[5pt]
 & - \frac{e^{\Phi/2}}{12} \Big(  F_{MPK\Lambda} F^{\;\; PK\Lambda}_N - \frac{3}{32} g_{MN} F^2_4 \Big) - \frac{e^{-\Phi}}{4} \Big(  H_{MPK} H^{\;\; PK}_N - \frac{1}{12} g_{MN} H^2_3 \Big) \, ,
 \\[5pt]
 \label{DilatonEq}
 0 & = \nabla^M \nabla_M \Phi - \frac{3}{8} e^{3\Phi/2} F^2_2 - \frac{e^{\Phi/2}}{96} F^2_4 + \frac{e^{-\Phi}}{12} H^2_3 \, ,
 \\[5pt]
 \label{Maxwell1}
 0 & = \nabla^M \big(  e^{-\Phi} H_{MNP} \big) - \frac{e^{\Phi/2}}{2} F_{NPK\Lambda} F^{K\Lambda} + \frac{1}{2 \cdot 4! \cdot 4!} \varepsilon_{M_1 \ldots M_8 NP} F^{M_1 \ldots M_4} F^{M_5 \ldots M_8} \, ,
 \\[5pt]
  \label{Maxwell2}
 0 & = \nabla^M \big(  e^{3\Phi/2} F_{MN}  \big) + \frac{e^{\Phi/2}}{6} F_{PK\Lambda N} H^{PK\Lambda} \, ,
 \\[5pt]
  \label{Maxwell3}
 0 & = \nabla^M \big(  e^{\Phi/2} F_{MNPK}  \big) - \frac{1}{144} \varepsilon_{M_1 \ldots M_7 NPK} F^{M_1 \ldots M_4} H^{M_5 \ldots M_7} \, ,
\end{align}
where $\varepsilon_{M_1 \ldots M_{10}}$ is the totally antisymmetric Levi-Civita \emph{tensor}.

An alternative way to present the dilaton equation \eqref{DilatonEq} is by eliminating its dependence in the RR fields. For this reason we first consider the trace of the Einstein equation \eqref{EinsteinEq}:
\begin{equation}
 0 = R - \frac{1}{2} \, \big(  \partial \Phi  \big)^2 - \frac{3}{16} \, e^{3\Phi/2} \, F^2_2 - \frac{e^{\Phi/2}}{192} \, F^2_4 - \frac{e^{-\Phi}}{24} \, H^2_3 \, .
\end{equation}
Using this we can re-write eq. \eqref{DilatonEq} as:
\begin{equation}
\label{DilatonEqNS}
 0 = -2 \, R + \nabla^M \nabla_M \Phi + \big(  \partial \Phi  \big)^2 + \frac{e^{-\Phi}}{6} \, H^2_3 \, .
\end{equation}

\subsection{Fluctuations of the equations of motion}

Here we derive the linearized equations of motion for the fluctuations of the supergravity fields. For this purpose we consider the following perturbation scheme:
\begin{equation}
 \begin{aligned}
  & g_{MN} = \bar{g}_{MN} + \d g_{MN} \, , \qquad \Phi = \bar{\Phi} + \varphi \, , \qquad H_3 = \bar{H}_3 + \d H_3 \, ,
  \\[5pt]
  & F_2 = \bar{F}_2 + \d F_2 \, , \qquad F_4 = \bar{F}_4 + \d F_4 \, ,
 \end{aligned}
\end{equation}
where we use the bar notation for the background values of the various fields. Notice that the background metric $\bar{g}_{MN}$ is the metric in the \emph{Einstein frame} which is conformally related to the metric $\tilde{g}_{MN}$ through a warp factor as in eq. \eqref{ConfResc}. Let us now continue with the fluctuations of the equations of motion.

\subsubsection{The Einstein equation}
\label{appEinsteinEq}

Before we start fluctuating the Einstein equation \eqref{EinsteinEq} we would like to re-write it in a more uniform way as:
\begin{equation}
 0 = R_{MN} - \frac{1}{2} \partial_M \Phi \, \partial_N \Phi - \frac{1}{2} \sum\limits_{p = 2}^{4} \g_p \, e^{\a_p \Phi} \Big[  \big(  \cA^2_p  \big)_{MN} - \b_p \, g_{MN} \cA^2_p  \Big] \, ,
\end{equation}
where we denote $\big(  \cA^2_p  \big)_{MN} := \cA_{M M_1 \ldots M_{p-1}} \,  \cA^{\;\;\; M_1 \ldots M_{p-1}}_N$ and we consider:
\begin{equation}
 \begin{aligned}
  \cA_p := \big(   F_2, H_3, F_4  \big) \, , \qquad \a_p := \Big(  \frac{3}{2}, -1, \frac{1}{2}  \Big) \, , \qquad \b_p := \Big(  \frac{1}{16}, \frac{1}{12}, \frac{3}{32}  \Big) \, , \qquad \g_p := \Big(  1, \frac{1}{2}, \frac{1}{6}  \Big) \, .
 \end{aligned}
\end{equation}
We also introduce the dot product for a tensor $\cA_p$ as:
\begin{equation}
 \begin{aligned}
  & \big(  \cA_p  \big)_M \cdot \big(  \bar{\cA}_p  \big)_N = \cA_{M \Sigma_1 \ldots \Sigma_{p-1}} \, \bar{\cA}^{\;\;\; \Sigma_1 \ldots \Sigma_{p-1}}_N \quad \& \quad \big(  \cA_p  \big) \cdot \big(  \bar{\cA}_p  \big) = \cA_{\Sigma_1 \ldots \Sigma_p} \, \bar{\cA}^{\Sigma_1 \ldots \Sigma_p} \, ,
  \\[5pt]
  & \big(  \cA_p  \big)_M \cdot \big(  \tilde{\cA}_p  \big)_N = \cA_{M \Sigma_1 \ldots \Sigma_{p-1}} \, \tilde{\cA}^{\;\;\; \Sigma_1 \ldots \Sigma_{p-1}}_N \quad \& \quad \big(  \cA_p  \big) \cdot \big(  \tilde{\cA}_p  \big) = \cA_{\Sigma_1 \ldots \Sigma_p} \, \tilde{\cA}^{\Sigma_1 \ldots \Sigma_p} \, ,
 \end{aligned}
\end{equation}
where we use the bar and tilde notation in order to stress that the indices are raised with the background metrics $\bar{g}_{MN}$ and $\tilde{g}_{MN}$ respectively.
After some algebra one ends up with:
\begin{equation}
 \begin{aligned}
  0 & = \frac{1}{2} \, \tilde{\nabla}^{\Sigma} \tilde{\nabla}_M h_{\Sigma N} + \frac{1}{2} \, \tilde{\nabla}^{\Sigma} \tilde{\nabla}_N h_{\Sigma M} - \frac{1}{2} \, \tilde{\nabla}^2 h_{MN} - \frac{1}{2} \, \tilde{\nabla}_N \tilde{\nabla}_M \tilde{h} + 4 \, \tilde{\nabla}^{\Sigma} A \, \tilde{\nabla}_M h_{\Sigma N}
  \\[5pt]
  & + 4 \, \tilde{\nabla}^{\Sigma} A \, \tilde{\nabla}_N h_{\Sigma M}
  - h_{MN} \,  \tilde{\nabla}^2 A - 8 \, h_{MN} \, \big(  \tilde{\nabla} A  \big)^2 - 4 \, \tilde{\nabla}^P A \tilde{\nabla}_P h_{MN}
  - \frac{1}{2} \, \partial_M \varphi \, \partial_N \bar{\Phi} - \frac{1}{2} \, \partial_M \bar{\Phi} \, \partial_N \varphi
  \\[5pt]
  & - \frac{1}{2} \sum\limits_{p = 2}^{4} \g_p \, e^{2 (1 - p) A + \a_p \bar{\Phi}} \Bigg[  \big(  \d\cA_p  \big)_M \cdot \big(  \tilde{\cA}_p  \big)_N
  + \big(  \tilde{\cA}_p  \big)_M \cdot \big(  \d\cA_p  \big)_N - \b_p \, h_{MN} \tilde{\cA}^2_p
  \\[5pt]
  & - (p - 1) \, h_{PK} \, \tilde{\cA}^{\qquad\quad\;\;\;\; P}_{M \Sigma_1 \ldots \Sigma_{p-2}} \, \tilde{\cA}^{\;\;\;\Sigma_1 \ldots \Sigma_{p-2} K}_{N}  \Bigg] - \frac{\varphi}{2} \, \sum\limits_{p = 2}^{4} \a_p \, \g_p \, e^{2 (1 - p) A + \a_p \bar{\Phi}} \big(  \tilde{\cA}^2_p  \big)_{MN} + \tilde{g}_{MN} \, t \, .
 \end{aligned}
\end{equation}
where $t$ is defined as:
\begin{equation}
 \begin{aligned}
  t & := \tilde{\nabla}^{\Sigma} h_{\Sigma P} \, \tilde{\nabla}^P A + h_{\Sigma P} \, \tilde{\nabla}^{\Sigma} \tilde{\nabla}^P A
  - \frac{1}{2} \, \, \tilde{\nabla}_{\Lambda} \tilde{h} \, \tilde{\nabla}^{\Lambda} A
  + 8 \, h_{P \Sigma} \, \tilde{\nabla}^P A \, \tilde{\nabla}^{\Sigma} A
  \\[5pt]
  & + \frac{\varphi}{2} \, \sum\limits_{p = 2}^{4} \a_p \, \b_p \, \g_p \, e^{2 (1 - p) A + \a_p \bar{\Phi}} \tilde{\cA}^2_p
  + \frac{1}{2} \sum\limits_{p = 2}^{4} \b_p \, \g_p \, e^{2 (1 - p) A + \a_p \bar{\Phi}} \Big[  2 \, \big(  \d\cA_p  \big) \cdot \big(  \tilde{\cA}_p  \big) 
  \\[5pt]
  & - p \, h_{PK} \, \tilde{\cA}^{\qquad\quad P}_{\Sigma_1 \ldots \Sigma_{p-1}} \, \tilde{\cA}^{\Sigma_1 \ldots \Sigma_{p-1} K}  \Big] \, .
 \end{aligned}
\end{equation}
Here by $\tilde{h}$ we mean the contraction $\tilde{h} := \tilde{g}^{MN} h_{MN}$

\subsubsection{The dilaton equation}

If we consider the fluctuation of the dilaton equation \eqref{DilatonEqNS} we arrive at the following result:
\begin{equation}
 \begin{aligned}
  0 & = 2 \, \tilde{R}^{MN} \,  h_{MN} - 2 \, \tilde{\nabla}^M \tilde{\nabla}^N h_{MN} + 2\, \tilde{\nabla}^2 \tilde{h} + \tilde{\nabla}^2 \varphi + 8 \, \tilde{\nabla}^M A \, \tilde{\nabla}_M \varphi - h_{MN} \, \tilde{\nabla}^M \tilde{\nabla}^N \bar{\Phi} 
  \\[5pt]
  & - \tilde{\nabla}^M \bar{\Phi} \, \tilde{\nabla}^N h_{MN} - 8 \, h_{MN} \, \tilde{\nabla}^M \bar{\Phi} \, \tilde{\nabla}^N A + \frac{1}{2} \, \tilde{\nabla}^M \bar{\Phi} \, \tilde{\nabla}_M \tilde{h} - h_{MN} \, \tilde{\nabla}^M \bar{\Phi} \, \tilde{\nabla}^N \bar{\Phi}+ 2 \, \tilde{\nabla}^M \bar{\Phi} \, \tilde{\nabla}_M \varphi 
  \\[5pt]
  & - 36 \,  h_{MN} \, \tilde{\nabla}^M \tilde{\nabla}^N A - 144 \,  h_{MN} \, \tilde{\nabla}^M A \, \tilde{\nabla}^N A - 36 \, \tilde{\nabla}^M A \, \tilde{\nabla}^N h_{MN} + 18 \, \tilde{\nabla}^M A \, \tilde{\nabla}_M \tilde{h} 
  \\[5pt]
  & - \frac{e^{- \bar{\Phi} - 4 A}}{6} \, \varphi \, \tilde{H}^2_3 +  \frac{e^{- \bar{\Phi} - 4 A}}{3} \, \big(  \d H_3  \big) \cdot \big(  \tilde{H}_3  \big) -  \frac{e^{- \bar{\Phi} - 4 A}}{2} \, h_{PK} \, \tilde{H}^{\quad \;\; P}_{MN} \, \tilde{H}^{MN K} \, .
 \end{aligned}
\end{equation}

\subsubsection{The Maxwell equations}

Let us now continue with the variation of the Maxwell equations \eqref{Maxwell1}, \eqref{Maxwell2} and \eqref{Maxwell3}. The results are summarized in the following three equations:
\vskip 5pt
\noindent\underline{\emph{The equation for the NS three-form:}}
\begin{equation}
 \begin{aligned}
 0 & =12 \; h_{MN} \tilde{\nabla}^M A \; \tilde{H}^{N}_{\;\; K_1 K_2} + h_{MN} \tilde{\nabla}^M \bar{\Phi} \; \tilde{H}^{N}_{\;\; K_1 K_2} + \varphi \; \tilde{\nabla}^M \bar{\Phi} \; \tilde{H}_{M K_1 K_2} - 4 \, \varphi \; \tilde{\nabla}^M A \; \tilde{H}_{M K_1 K_2}
   \\[5pt]
   & - \tilde{\nabla}^M \varphi \; \tilde{H}_{M K_1 K_2} + 4 \; \tilde{\nabla}^M A \; \d H_{M K_1 K_2} - \tilde{\nabla}^M \bar{\Phi} \; \d H_{M K_1 K_2} - \varphi \; \tilde{\nabla}^M \tilde{H}_{M K_1 K_2} + \tilde{\nabla}^M \d H_{M K_1 K_2}
  \\[5pt]
  & - \frac{1}{2} \big(   2 \; \tilde{\nabla}^M h_{\Sigma M} - \tilde{\nabla}_{\Sigma} \tilde{h} \big) \, \tilde{H}^{\Sigma}_{\;\;\; K_1 K_2} - h_{MN} \, \tilde{\nabla}^N \tilde{H}^M_{\;\;\; K_1 K_2}  - \tilde{\nabla}_N h_{\Sigma K_1} \, \tilde{H}^{N \Sigma}_{\;\;\;\;\;\; K_2} + \tilde{\nabla}_N h_{\Sigma K_2} \, \tilde{H}^{N \Sigma}_{\;\;\;\;\;\; K_1}
  \\[5pt]
  & - \frac{e^{\frac{3}{2} \bar{\Phi} - 2 A}}{4} \varphi \; \tilde{F}_{K_1 K_2 K\Lambda} \tilde{F}^{K\Lambda} - \frac{e^{\frac{3}{2} \bar{\Phi} - 2 A}}{2} \d \tilde{F}_{K_1 K_2 K\Lambda} \tilde{F}^{K\Lambda}
  - \frac{e^{\frac{3}{2} \bar{\Phi} - 2 A}}{2} \tilde{F}^{\;\;\;\;\;\;\; K\Lambda}_{K_1 K_2} \big(  \d \tilde{F}_{K \Lambda} - 2 \, h_{K M}  \tilde{F}^{M}_{\;\;\;\; \Lambda}  \big) 
  \\[5pt]
  & + \frac{e^{\bar{\Phi} - 4 A}}{2 \cdot 4! \cdot 4!} \, \Big[  \frac{1}{2} \, \tilde{h} \, \tilde{\varepsilon}_{M_1 \ldots M_8 K_1 K_2}  \tilde{F}^{M_1 \ldots M_4} \tilde{F}^{M_5 \ldots M_8}
  + 2 \, \tilde{\varepsilon}^{M_1 \ldots M_4}_{\qquad\;\;\; M_5 \ldots M_8 K_1 K_2} \d F_{M_1 \ldots M_4} \tilde{F}^{M_5 \ldots M_8} 
  \\[5pt]
  & - 8 \, h_{M_1 M'_1} \, \tilde{\varepsilon}^{M_1}_{ \;\;\;\; M_2 \ldots M_8 K_1 K_2}  \tilde{F}^{M'_1 M_2 M_3 M_4} \tilde{F}^{M_5 \ldots M_8} \Big] \, .
 \end{aligned}
\end{equation}
\vskip 5pt
\noindent\underline{\emph{The equation for the RR two-form:}}
\begin{equation}
 \begin{aligned}
0 & =  12 \; h_{MN} \tilde{\nabla}^M A \; \tilde{F}^{N}_{\;\; K_1} - \frac{3}{2} \; h_{MN} \tilde{\nabla}^M \bar{\Phi} \; \tilde{F}^{N}_{\;\; K_1} + \frac{9}{4} \; \varphi \; \tilde{\nabla}^M \bar{\Phi} \; \tilde{F}_{M K_1} + 9 \; \varphi \; \tilde{\nabla}^M A \; \tilde{F}_{M K_1}
   +  \frac{3}{2} \; \tilde{\nabla}^M \varphi \; \tilde{F}_{M K_1}
   \\[5pt]
   & + 6 \; \tilde{\nabla}^M A \; \d F_{M K_1} 
   + \frac{3}{2} \; \tilde{\nabla}^M \bar{\Phi} \; \d F_{M K_1}
   +  \frac{3}{2} \varphi \; \tilde{\nabla}^M \tilde{F}_{M K_1}
   + \tilde{\nabla}^M \d F_{M K_1}
   - h_{MN} \, \tilde{\nabla}^N \tilde{F}^M_{\;\;\; K_1} 
   - \tilde{\nabla}_N h_{\Sigma K_1} \, \tilde{F}^{N \Sigma}
   \\[5pt]
   & - \frac{1}{2} \big(   2 \; \tilde{\nabla}^M h_{\Sigma M} - \tilde{\nabla}_{\Sigma} \tilde{h} \big) \, \tilde{F}^{\Sigma}_{\;\;\; K_1} + \frac{e^{- 4 A - \bar{\Phi}}}{12} \varphi \tilde{F}_{PK\Lambda K_1} \tilde{H}^{PK\Lambda} 
   + \frac{e^{- 4 A - \bar{\Phi}}}{6} \d F_{PK\Lambda K_1} \tilde{H}^{PK\Lambda}
   \\[5pt]
   & + \frac{e^{- 4 A - \bar{\Phi}}}{6} \tilde{F}^{PK\Lambda}_{\quad\;\;\; K_1} \big(   \d H_{PK\Lambda} - 3 \, h_{P \Sigma} \, \tilde{H}^{\Sigma}_{\;\;\; K \Lambda}   \big) \, .
 \end{aligned}
\end{equation}
\vskip 5pt
\noindent\underline{\emph{The equation for the RR four-form:}}
\begin{equation}
 \begin{aligned}
  0 & = 12 \; h_{MN} \tilde{\nabla}^M A \; \tilde{F}^{N}_{\;\;\; K_1 K_2 K_3} - \frac{1}{2} \; h_{MN} \tilde{\nabla}^M \bar{\Phi} \; \tilde{F}^{N}_{\;\;\; K_1 K_2 K_3} + \frac{\varphi}{4} \; \tilde{\nabla}^M \bar{\Phi} \; \tilde{F}_{M K_1 K_2 K_3} + \varphi \; \tilde{\nabla}^M A \; \tilde{F}_{M K_1 K_2 K_3}
   \\[5pt]
   & +  \frac{1}{2} \; \tilde{\nabla}^M \varphi \; \tilde{F}_{M K_1 K_2 K_3}
    + \frac{\varphi}{2} \; \tilde{\nabla}^M \tilde{F}_{M K_1 K_2 K_3}
    + 2 \; \tilde{\nabla}^M A \; \d F_{M K_1 K_2 K_3}
    + \frac{1}{2} \; \tilde{\nabla}^M \bar{\Phi} \; \d F_{M K_1 K_2 K_3}
   \\[5pt]
   & - \frac{1}{2} \big(   2 \; \tilde{\nabla}^M h_{\Sigma M} - \tilde{\nabla}_{\Sigma} \tilde{h} \big) \, \tilde{H}^{\Sigma}_{\;\;\; K_1 K_2} - h_{MN} \, \tilde{\nabla}^N \tilde{F}^M_{\;\;\; K_1 K_2 K_3}  - \frac{1}{2} \big(   2 \; \tilde{\nabla}^M h_{\Sigma M} - \tilde{\nabla}_{\Sigma} \tilde{h} \big) \, \tilde{F}^{\Sigma}_{\;\;\; K_1 K_2 K_3}
   \\[5pt]
   & - \tilde{\nabla}_N h_{\Sigma K_1} \, \tilde{F}^{N \Sigma}_{\;\;\;\;\;\; K_2 K_3} + \tilde{\nabla}_N h_{\Sigma K_2} \, \tilde{F}^{N \Sigma}_{\;\;\;\;\;\; K_1 K_3} - \tilde{\nabla}_N h_{\Sigma K_3} \, \tilde{F}^{N \Sigma}_{\;\;\;\;\;\; K_1 K_2}
   \\[5pt]
   & - \frac{e^{- 2 A - \frac{\bar{\Phi}}{2}}}{144} \, \Big[   \frac{1}{2} \, \tilde{h} \, \tilde{\varepsilon}_{M_1 \ldots M_7 K_1 K_2 K_3} \;  \tilde{F}^{M_1 \ldots M_4} \; \tilde{H}^{M_5 M_6 M_7}
   + \tilde{\varepsilon}^{M_1 \ldots M_4}_{\qquad\;\;\; M_5 \ldots M_7 K_1 K_2 K_3} \; \d F_{M_1 \ldots M_4} \; \tilde{H}^{M_5 \ldots M_7}
   \\[5pt]
   & - 4 \, h_{M_1 M'_1} \, \tilde{\varepsilon}^{M_1}_{ \;\;\;\; M_2 \ldots M_7 K_1 K_2 K_3} \;  \tilde{F}^{M'_1 M_2 \ldots M_4} \; \tilde{H}^{M_5 M_6 M_7}
  + \tilde{\varepsilon}^{M_1 M_2 M_3}_{\qquad\;\;\;\;\;\; M_4 \ldots M_7 K_1 K_2 K_3} \; \d H_{M_1 M_2 M_3} \; \tilde{F}^{M_4 \ldots M_7}
  \\[5pt]
  & - 3 \, h_{M_1 M'_1} \, \tilde{\varepsilon}^{M_1}_{ \;\;\;\; M_2 \ldots M_7 K_1 K_2 K_3} \;  \tilde{H}^{M'_1 M_2 M_3} \;  \tilde{F}^{M_4 \ldots M_7}  \Big] \, .
 \end{aligned}
\end{equation}

\section{WKB approximation}

\label{WKBmethod}

In this appendix we review the WKB approximation method following the lines of \cite{Russo:1998by}. The formalism that was developed in \cite{Russo:1998by} applies to eigenvalue problems for second order differential equations of the following form:
\begin{equation}
\label{WKBDE}
 \partial_r \big(   p(r) \partial_r \Psi  \big) + \Big(  M^2 w(r) + q(r)  \Big) \Psi = 0 \, ,
\end{equation}
where $M$ represents the eigenvalue (in our case it represents the graviton mass). The functions $p(r), w(r)$ and $q(r)$ are independent of $M$. When eq. \eqref{WKBDE} is written in appropriate variables there is a point $r_*$ where the above functions behave as:
\begin{equation}
 p \approx p_1 (r - r_*)^{s_1} \, , \qquad w \approx w_1 (r - r_*)^{s_2} \, , \qquad q \approx q_1 (r - r_*)^{s_3} \qquad \textrm{as} \quad r \rightarrow r_* \, ,
\end{equation}
where $p_1, w_1, q_1$ and $s_1, s_2, s_3$ are constants. Similarly, we assume
\begin{equation}
 p \approx p_2 r^{t_1} \, , \qquad w \approx w_2 r^{t_2} \, , \qquad q \approx q_2 r^{t_3} \qquad \textrm{as} \quad r \rightarrow \infty \, ,
\end{equation}
where again $p_2, w_2, q_2$ and $t_1, t_2, t_3$ are constants.

Using the above expansions  one can derive an approximate formula for the eigenvalue $M$ whose accuracy is good for large enough values of $M$. This formula reads:
\begin{equation}
 M^2 = \frac{\pi^2}{\xi^2} \, \nu \Big(  \nu - 1 + \frac{\a_2}{\a_1} + \frac{\b_2}{\b_1} \Big) + \cO(\nu^0) \, , \qquad \nu = 1, 2, \ldots \, . \, ,
\end{equation}
where $\xi$ is given by:
\begin{equation}
 \xi := \int\limits^{\infty}_{r_*} dr \sqrt{\frac{w}{p}} \, .
\end{equation}
Also the constants $\a_1, \a_2$ and $\b_1, \b_2$ are determined by the expansion parameters:
\begin{equation}
 \a_1 = s_2 - s_1 + 2 \, , \qquad \b_1 = t_1 - t_2 - 2
\end{equation}
and
\begin{equation}
 \begin{aligned}
  & \a_2 = |s_1 - 1| \qquad \textrm{or} \qquad \a_2 = \sqrt{(s_1 - 1)^2 - 4 \frac{q_1}{p_1}} \quad (\textrm{if} \quad s_3 - s_1 + 2 = 0) \, ,
  \\[5pt]
  & \b_2 = |t_1 - 1| \qquad \textrm{or} \qquad \b_2 = \sqrt{(t_1 - 1)^2 - 4 \frac{q_2}{p_2}} \quad (\textrm{if} \quad t_1 - t_3 - 2 = 0) \, .
 \end{aligned}
\end{equation}



\end{document}